\documentclass{aa}
\usepackage{graphicx}
\usepackage{txfonts}
\usepackage{multirow}
\newcommand{\zw}{I~Zw~1}
\newcommand{\dif}{\mathrm{d}}
\usepackage{upgreek}
\begin{document} 

   \title{The outflowing ionised gas of I~Zw~1 observed by HST COS\thanks{The stacked HST COS spectrum is available in electronic form at the CDS via anonymous ftp to cdsarc.u-strasbg.fr (130.79.128.5) or via http://cdsweb.u-strasbg.fr/cgi-bin/qcat?J/A+A.}}

   \author{A. Jur\'{a}\v{n}ov\'{a}\inst{1,2},
          E. Costantini\inst{1,2},  
          G. A. Kriss\inst{3}, 
          M. Mehdipour\inst{3},
          W. N. Brandt\inst{4,5,6},
          L. Di Gesu\inst{7},
          A. C. Fabian\inst{8},
          L. Gallo\inst{9},
          M. Giustini\inst{10},
          D. Rogantini\inst{11}, \and
          D. R. Wilkins\inst{12} 
          }

   \institute{SRON Netherlands Institute for Space Research, Niels Bohrweg 4, NL-2333 CA Leiden, the Netherlands\\
              \email{a.juranova@sron.nl}
    \and Anton Pannekoek Institute, University of Amsterdam, Postbus 94249, NL-1090 GE Amsterdam, the Netherlands
    \and Space Telescope Science Institute, 3700 San Martin Drive, Baltimore, MD 21218, USA
    \and Department of Astronomy and Astrophysics, 525 Davey Lab, The Pennsylvania State University, University Park, PA 16802, USA
    \and Institute for Gravitation and the Cosmos, The Pennsylvania State University, University Park, PA 16802, USA
    \and Department of Physics, 104 Davey Laboratory, The Pennsylvania State University, University Park, PA 16802, USA
    \and Agenzia Spaziale Italiana, Via del Politecnico snc, I-00133 Roma, Italy
    \and Institute of Astronomy, University of Cambridge, Madingley Road, Cambridge CB3 0HA, UK
    \and Department of Astronomy and Physics, Saint Mary’s University, 923 Robie Street, Halifax, NS B3H 3C3, Canada
    \and Centro de Astrobiolog\'{i}a, CSIC-INTA, Camino Bajo del Castillo s/n, Villanueva de la Cañada, 28692 Madrid, Spain
    \and Department of Astronomy and Astrophysics, University of Chicago, Chicago, IL 60637, USA
    \and Kavli Institute for Particle Astrophysics and Cosmology, Stanford University, 452 Lomita Mall, Stanford, CA 94305, USA
             }

    \date{Received February 8, 2024; accepted March 20, 2024}
 
  \abstract
  % context heading (optional)
   {}
  % aims heading (mandatory)
   {We present an analysis of the Hubble Space Telescope Cosmic Origins Spectrograph spectrum of I~Zw~1 aiming to probe the absorbing medium associated with the active galactic nucleus (AGN).} 
  % methods heading (mandatory)
   {We fitted the emission spectrum and performed spectral analysis of the identified absorption features to derive the corresponding ionic column densities and covering fractions of the associated outflows. We employed photoionisation modelling to constrain the total column density and the ionisation parameter of four detected kinematic components. By investigating the implications of the results together with the observed kinematic properties of both emission and absorption features, we derived constraints on the structure and geometry of the absorbing medium in the AGN environment.}
  % results heading (mandatory)
   {We find and characterise absorption line systems from outflowing ionised gas in four distinct kinematic components, located at $-60$, $-280$, $-1950$, and $-2900~\mathrm{km\,s^{-1}}$ with respect to the source rest frame. While the two slower outflows are consistent with a full covering of the underlying radiation source, the well-constrained doublet line ratios of the faster two, higher column density, outflows suggest partial covering, with a covering fraction of $C_\mathrm{f}\sim 0.4$. The faster outflows show also line-locking in the \ion{N}{V} doublet, a signature of acceleration via line absorption. This makes I~Zw~1 possibly the closest object that shows evidence for hosting line-driven winds. The observed $-1950~\mathrm{km\,s^{-1}}$ absorption is likely due to the same gas as an X-ray warm absorber. Furthermore, the behaviour in UV and X-ray bands implies that this outflow has a clumpy structure. We find that the highly asymmetric broad emission lines in I~Zw~1, indicative of a collimated, outflowing broad line region, are covered by the absorbing gas. Finally, the strongest UV--X-ray absorber may be connected to some of the blueshifted line emission, indicative of a more spatially extended structure of this ionised medium.}
  % conclusions heading (optional)
   {}
    \authorrunning{Jur\'{a}\v{n}ov\'{a} et al.}
   \maketitle
   
%-------------------------------------------------------------------
\section{Introduction}\label{sec:intro}

Ultraviolet and X-ray observations of active galactic nuclei (AGNs) from the past decades suggest that ionised outflows are nearly universally present in these objects \citep[e.g.][]{Crenshaw2003a, Laha2021}. This tenuous gas, photoionised by the strong central radiation field, can be revealed via line absorption, typically blueshifted by several hundred to tens of thousands $\rm km~s^{-1}$ with respect to the source rest frame. When observed in the X-ray band, outflows with moderate velocities and the ionisation parameter typically in the range of $\log \xi \sim 0\!-\!3$, where $\xi$ is in units of $\rm erg~s^{-1}~cm $ (see eq. \ref{eq:xi}), are referred to as warm absorbers \citep[WA;][]{Reynolds1995, Tombesi2013, Laha2014}. The associated UV absorption in resonance lines of hydrogen (\ion{Ly}{$\alpha$}), \ion{C}{iv}, and a few other ions has been studied in great detail \citep[see e.g.][and the references therein]{Crenshaw2003b, Gibson2009}. However, despite the frequent detection of ionised outflows in high-resolution spectra, the physical properties, impact, and circumstances of their origin are not yet fully understood \citep[see e.g.][for a review]{Giustini2019}.

While velocity-dependent properties of these relatively highly ionised outflows can be identified in high-resolution UV spectra, the absorbers are typically detected only in a few strong transitions. In contrast, the X-ray band is richer in absorption lines, but due to instrumental limitations, physical characterisation of the gas is usually limited to the total column density, ionisation, and the mean radial velocity. Observations in both the UV and X-ray bands can, however, provide further insight into the structure of the gas \citep{Costantini2010, Laha2021}. In some cases, the absorption features show a clear connection to a gas of the same properties \citep[e.g.][]{Mathur1995, Arav2020}, but discrepancies between the UV and X-ray absorption can allude to a more complex geometry and behaviour of these outflows \citep[e.g.][]{Ebrero2013}.

The outward motion of the ionised material, reached despite the gravitational pull of the central supermassive black hole, can be achieved via several processes. Thermal pressure can launch winds from large distances from the central engine (e.g. from the broad line region or dusty torus) with velocities of several hundreds to low thousands of $\rm km~s^{-1}$ \citep[e.g.][]{Begelman1983, KrolikKriss2001, Everett2007, Proga2007, Mizumoto2019a}. Radiation-pressure driving can overcome the gravitational pull of the supermassive black hole at smaller distances, provided a sufficiently high flux and favourable spectral distribution of the incident radiation, dominant in the UV band \citep[e.g.][]{Mushotzky1972, Scargle1973, Proga2004}. On larger scales, beyond the dust sublimation radius, radiation pressure is expected to be an important factor again due to the relatively large cross-section of dust grains \citep{Fabian2006, Fabian2008}. Finally, outflows can be accelerated also in magnetic fields via magneto-centrifugal and magnetic pressure forces \citep[e.g.][]{Blandford1982, Contopoulos1994, Konigl1994, Fukumura2010}. 

A phenomenon known as line-locking \citep{Milne1926}, commonly observed in quasar spectra \citep[e.g.][]{Ganguly2003, Mas-Ribas2019, Veilleux2022}, is believed to distinctly identify ongoing radiative driving. This effect is observed in sources with at least two kinematic components and is manifested by an overlap of absorption doublet components in the spectrum. This alignment is interpreted as evidence for a dynamically important momentum transfer via those transitions.

In the geometrical configuration, the slower (shielding) component lies between the emission source and the other, faster, outflow, which also experiences greater total outward acceleration. When the outer outflow reaches a velocity equal to the doublet separation, it becomes effectively shielded by the slower outflow, as the flux at the corresponding wavelength suddenly decreases. This prevents the shielded cloud from being accelerated further and a line-locked position in velocity space is established. This velocity separation will be maintained regardless of the total acceleration of the shielding cloud for as long as the acceleration of the shielded one via the red doublet component stays equal to, or greater than, what the acceleration difference between the two outflows would be outside of the shadow.

This alignment could also be established by chance, but the prevalence of line-locked absorbing systems, seen in as many as two-thirds of SDSS quasars with multiple \ion{C}{iv} absorbers \citep{Bowler2014} or multiple line-locked doublets in one quasar \citep{Srianand2002}, clearly support line-locking as a natural consequence of radiative acceleration. 

In this work, we aim at characterising the ionised outflows in \object{\zw{}}. This object has prototypical optical properties of a narrow-line Seyfert 1 \citep[][]{Osterbrock1985, Sargent1968, Phillips1976, Gallo2018}, and for its high luminosity, it is also considered a quasar \citep{Schmidt1983}. The AGN is hosted in a spiral galaxy, which is undergoing a nuclear starburst \citep[e.g.][]{Eckart1994, Fei2023}. The central supermassive black hole has a mass of $9\times 10^{6} M_{\odot}$ \citep{Huang2019}, with an accretion rate close to the Eddington limit \citep{Wilkins2021}. The UV and optical spectrum of this AGN is rich in emission lines, including a plethora of \ion{Fe}{ii} and \ion{Fe}{iii} transitions \citep[][]{Sargent1968, BorosonGreen1992, Vestergaard2001, Veron-Cetty2004}. Interestingly, the emission-line centroids are generally not positioned at the source cosmological redshift but are shifted blueward, with the magnitude of this shift growing with the ionisation of the line-emitting gas, as first reported by \citet{Phillips1976}. Additionally, more highly-ionised lines were found to show stronger asymmetry, most prominent in the UV \ion{C}{iv} doublet, which shows only little emission in the red wing in contrast with a strong blue tail \citep{Laor1997}.

Analysing the UV spectrum of this source with the Faint Object Spectrograph (FOS) onboard the \textit{Hubble Space Telescope} (HST), \citet{Laor1997} identified an ionised absorber blueshifted to $\sim$$-1870~\rm km\,s^{-1}$ with respect to the source rest frame in the \ion{N}{v} and \ion{C}{iv} lines. \ion{O}{vi} absorption associated with two kinematic components was later detected in Far Ultraviolet Spectroscopic Explorer (FUSE) data by \citet[][and priv. comm.]{Kriss2002}. In the X-ray band, a low-ionisation intrinsic absorber has been first noted by \citet{Leighly1999b} in moderate-resolution \textit{ASCA} data. In the following campaigns with \textit{XMM-Newton}, this detection was confirmed by \citet{Costantini2007}. The high-resolution RGS data revealed that the WA properties were consistent with the UV absorption signature and that also more highly ionised components were present in the system. Additionally, their and subsequent observations proved that the WAs display complex variability \citep{Silva2018, Rogantini2022}.
 
Following up on these results, in this study we present simultaneous UV and X-ray observations of \zw{}, with the main focus on the ionised gas signatures seen in the high-resolution HST COS spectrum. 
In Section \ref{sec:observations}, we summarise the HST observation details, and we present the analysis of the spectrum and its outcomes in Section \ref{sec:analysis-results}. In particular, we address the spectral composition of the reconstructed incident ionising radiation, spectral fitting, and subsequent photoionisation modelling of the outflows. The discussion of the results is given in Section \ref{sec:discussion}, with a focus on the observed line-locking, connection to ionised absorption in the X-ray band, possible dust extinction effects, and any emission signatures related to the absorbing gas. Finally, we summarise our findings in Section \ref{sec:conclusions}. 

Throughout the paper, we adopt the following cosmological parameters: $H_0 = 70 \rm ~km ~s^{-1}~ Mpc^{-1}$, $\Omega_{\rm m} = 0.3$, and $\Omega_{\Lambda} = 0.7$. We use $\chi^2$-statistic for the UV data analysis and $C$-statistic \citep{Cash1979, Kaastra2017} for the X-ray spectral fitting. The reported uncertainties are calculated at 1$\sigma$ significance unless stated otherwise.

%--------------------------------------------------------------------
\section{Observations}\label{sec:observations}

\zw{} was observed by HST during eight orbits between 20 and 22 January 2015 (program 13811), simultaneously with two \textit{XMM-Newton} \citep{XMM} observations. For the results of the X-ray timing analysis, we refer the reader to \citet{Wilkins2017}. \citet{Silva2018} details the results from the high-resolution X-ray observations performed with the RGS. Here, we report on the new high-resolution UV observations of \zw{} performed with the HST Cosmic Origins Spectrograph (COS). 

\begin{figure*}
    \centering
    \resizebox{\hsize}{!}{\includegraphics{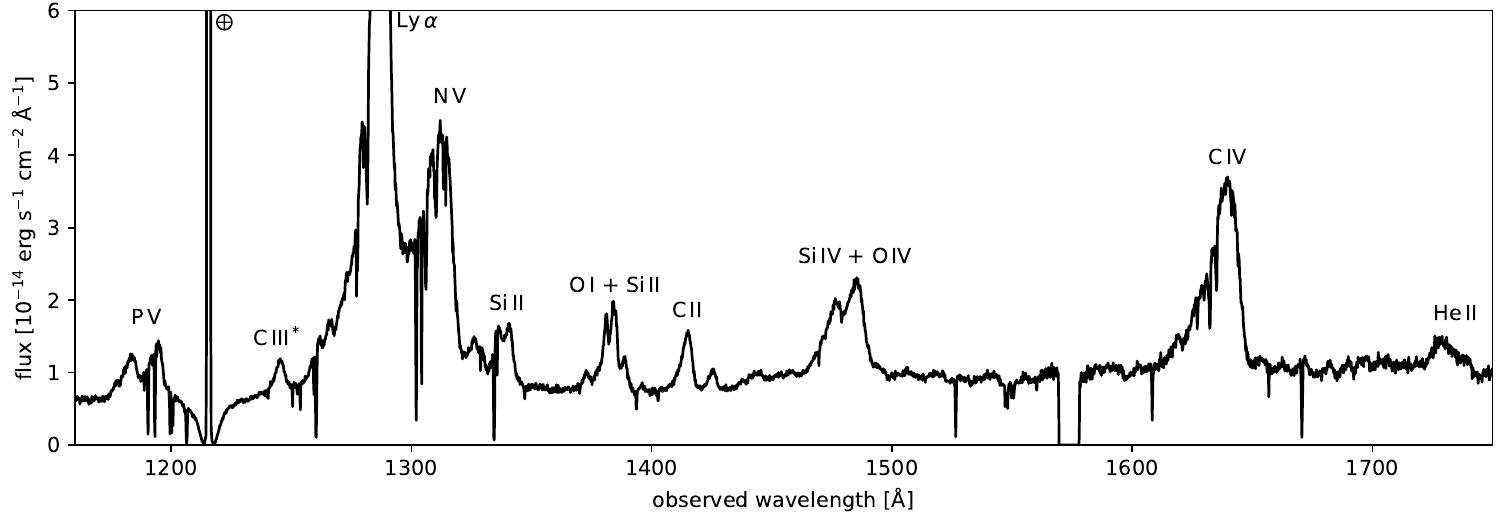}}
    \caption{Calibrated COS spectrum. The most prominent emission features are labelled, including the Geocoronal line in the centre of the Galactic \ion{Ly}{$\alpha$} absorption trough (Earth symbol). The data are binned  by a factor of 16 for clarity.}
    \label{fig:COS}
\end{figure*}

The COS spectra were taken with the Primary Science Aperture and G130M and G160M gratings, covering the far-UV spectrum in the range of 1133--1760\,\AA{} at a resolving power $\Delta\lambda/\lambda \sim 15\,000$--22\,000 \citep{HST-COS}. The observation details are listed in Table \ref{tab:COS}. The spectra were processed with the COS calibration pipeline version 3.0. The resulting stacked spectrum is plotted in Fig. \ref{fig:COS}.

\begin{table}[]
    \centering
    \caption{Observation details of the HST-COS 2015 observation campaign of \zw{}.}
    \begin{tabular}{lcccccc}
        \hline\hline
        Data set & Grating/tilt & Start time & Exposure \\
                    &              &   GMT &    s      \\
        \hline
        lckr01010 & G130M/1291 & Jan-20~08:14:22 & 960 \\
        lckr01020 & G130M/1327 & Jan-20~08:33:49 & 950 \\
        lckr01030 & G130M/1309 & Jan-20~09:34:35 & 1900 \\
        lckr01040 & G160M/1577 & Jan-20~10:12:22 & 3000 \\
        lckr01050 & G160M/1589 & Jan-20~14:43:55 & 2560 \\
        lckr02010 & G130M/1291 & Jan-21~22:26:18 & 960 \\
        lckr02020 & G130M/1327 & Jan-21~22:45:45 & 950 \\
        lckr02030 & G130M/1309 & Jan-21~23:46:39 & 1900 \\
        lckr02040 & G160M/1577 & Jan-22~00:24:26 & 3000 \\
        lckr02050 & G160M/1589 & Jan-22~02:57:39 & 2560 \\
                \hline
    \end{tabular}
    \label{tab:COS}
\end{table}

%-------------------------------------------------------------------
\section{Analysis and results}\label{sec:analysis-results}

For the characterisation of the absorbers detected in the COS spectrum, it is necessary to construct the spectral energy distribution (SED) of the incident ionising radiation. This step is addressed in the text below, followed by the description of the COS spectral fitting with a focus on the emission lines and the imprinted absorption properties. The derivation of the ionic column densities from the observed absorption features concludes this section.

The results are derived assuming elemental abundances of \citet{Lodders2009} and solar metallicity. We adopt $z = 0.061169 $ extracted from the \ion{H}{i} 21 cm line \citep{Springob2005} for the systemic redshift of \zw{} throughout this paper.

\subsection{The ionising SED}\label{ssec:SED}

To describe the far-UV part of the SED, we extracted the continuum fluxes from six narrow parts of the COS spectrum largely unaffected by line emission. Additionally, to constrain the low energy end of the AGN spectrum, we took advantage of the simultaneous observations with the \textit{XMM-Newton} Optical Monitor \citep[OM;][]{OM}, taken with the six photometric filters onboard, i.e. \textit{UVW2}, \textit{UVM2}, \textit{UVW1}, \textit{U}, \textit{B}, and \textit{V}. We fitted these data in \textsc{SPEX} v3.07.03 \citep{SPEX, SPEX30703} using a spectral model composed as follows.

For the stellar light enhancing the optical flux, we adopted the bulge template of \citet{Kinney1996}, assuming the majority of the stellar emission in the spectrum comes from the central region of the host galaxy. We adopted a disc black-body component (\texttt{dbb} in SPEX) to describe the AGN continuum, with the electron temperature $T_{\rm dbb}$ fixed to 10 eV. To account for the Galactic dust extinction, we used the extinction law of \citet{Cardelli1989}, with $R_V = 3.1$ and a colour excess fixed to $E(B-V) = 0.057$ \citep{Schlafly2011}.

\begin{figure}
    \centering
    \resizebox{\hsize}{!}{\includegraphics{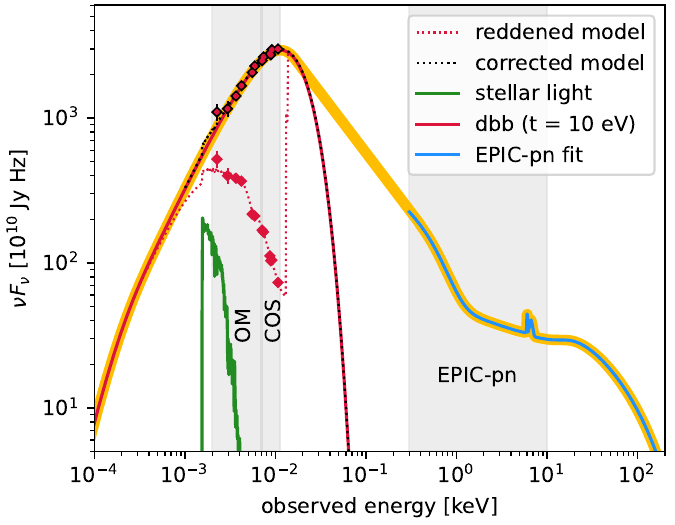}}
    \caption{Spectral energy distribution of \zw{} derived from the 2015 broadband observations. The best-fitting model (dotted red line) of the UV and optical data points (red diamonds) consists of a disc black body (solid red line), a stellar light component (green line), and a custom extinction curve. The reddening-corrected data and model are shown as well. The disc black body component is connected to the continuum model of the EPIC-pn spectrum \citep[][blue line]{Rogantini2022} to form the SED (yellow curve). The observationally covered bands are depicted with the grey-shaded areas.}
    \label{fig:SED}
\end{figure}

\begin{figure}
    \centering
    \resizebox{\hsize}{!}{\includegraphics{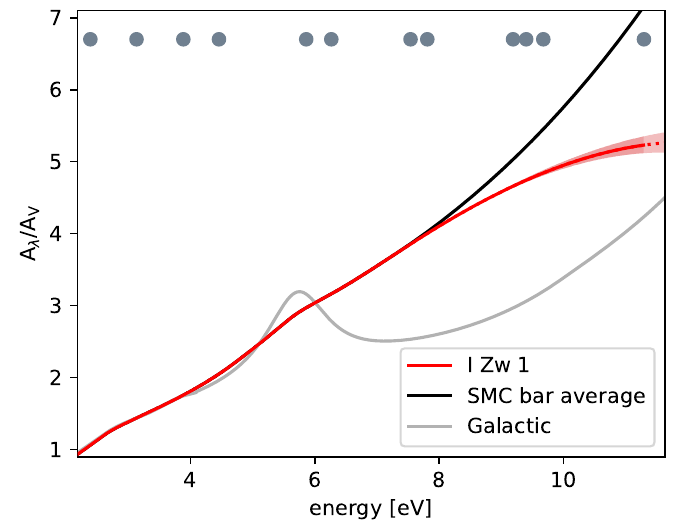}}
    \caption{Extinction in \zw{} derived from the COS and OM 2015 data. The extinction curve (red line) is based on the \citet{Gordon2003} SMC bar average curve (black). The relatively smaller amount of reddening above 8~eV is described by an additional parameter controlling the far-UV curvature, constrained from the fit to $c_4 = -0.34\pm 0.05$ (see the text for more details). This $1\sigma$ uncertainty is visualised with the shaded area surrounding the curve. The \citet{Fitzpatrick1999} Galactic extinction curve (grey) is given for reference. The positions of the data points used for the fitting are depicted with grey dots on the top.}
    \label{fig:extlaw}
\end{figure}

After the correction for the column of Galactic dust was applied, we still observed a steep blue-ward decrease of continuum flux from the optical to the UV, as can be seen in Fig. \ref{fig:SED}. Such a behaviour is indicative of substantial reddening, associated with the host. To account for this additional extinction, we used the \citet{Gordon2003} Small Magellanic Cloud (SMC) bar average model with $R_V = 3.1$, using the \citet{Fitzpatrick1999} parametrisation. However, a custom extinction law was necessary to obtain a good fit of the data. In particular, flattenning -- relative to the SMC extinction curve -- above 8 eV (1550~\AA) is needed (see Fig. \ref{fig:extlaw}), suggestive of a relative lack of grains with the corresponding size of $ \lesssim\!0.25~\mathrm{\mu m} $. We allowed two parameters to vary in the fit, namely the far-UV curvature parameter $c_4$ \citep[detailed in][]{Fitzpatrick2005}, constrained to $-0.34\pm 0.05$, and the colour excess, yielding $E(B-V) = 0.206\pm 0.006$. The best-fitting extinction curve is plotted in Fig. \ref{fig:extlaw} along with the \citet{Gordon2003} SMC bar average and \citet{Cardelli1989} Galactic extinction curves, for reference.

We note, however, that the lack of data beyond 11.3~eV prevents placing independent constraints on the extinction correction and the \texttt{dbb} parameters. To test the dependence of the atypical dust extinction law parameters on the assumed disc temperature, we fitted the data points while varying $T_{\rm dbb}$ over a physically meaningful range of values. Interestingly, we find that the shape of the extinction law in the observed energy range exhibits minimal sensitivity (within $2\sigma$) to the choice of $T_{\rm dbb}$. For instance, a substantially higher disc temperature $T_{\rm dbb} = 20~\rm eV$ yields $c_4 = -0.28\pm 0.05$.

Having the low energy end of the SED described, we composed the SED by joining the \texttt{dbb} profile with the X-ray model of \citet{Rogantini2022}, fitted to the concurrent \textit{XMM-Newton} EPIC-pn spectrum. The UV--X-ray gap was bridged by finding a common tangent of the \texttt{dbb} and the modelled X-ray soft excess at $\sim$$0.3~\rm keV$, dominated by a Comptonisation component. The resulting model is displayed in Fig. \ref{fig:SED}. The bolometric luminosity obtained from the SED model is $L_\mathrm{bol} = 10^{45.86}\,\rm erg\,s^{-1}$ and the ionising luminosity (defined as the luminosity between 1 and 1000 Ryd) is $L_\mathrm{ion} = 10^{45.51}\,\rm erg\,s^{-1}$.

\subsection{Underlying emission spectrum}\label{ssec:emission}

We modelled the COS spectrum around the observed or expected absorption features with a phenomenological model to obtain a good representation of the underlying emission. The fit was performed on the Galactic extinction-corrected spectrum. Considering the rather strong continuum reddening described in Sect. \ref{ssec:SED}, the true line emission might be significantly stronger. 

The emission lines peak blue-ward of the predicted positions obtained assuming the cosmological redshift ($ z = 0.061169 $). This shift is not uniform, as the lower ionisation lines, such as \ion{O}{i}, \ion{Si}{ii} or \ion{S}{ii} peak at $z = 0.0607$ (which corresponds to a velocity shift of $-130\, \rm km\,s^{-1}$), while higher ionisation lines such as \ion{P}{v}, \ion{N}{v} or \ion{C}{iv} peak at a lower redshift still, at $z = 0.05870$ (or $-700\, \rm km\,s^{-1}$). \ion{Ly}{$\alpha$}, produced in a broad range of ionisation, peaks at $z = 0.05987$. This confirms the trend observed in HST-FOS data by \citet{Laor1997}.

We modelled the emission spectrum with a power law for the continuum, and for the emission lines, we used Gaussian profiles. As most of the lines are strongly asymmetric (see Fig. \ref{fig:COS}), we allowed for different broadening of the Gaussians in the red and blue wings. We characterised this with an additional parameter $\gamma$, defined as $\rm FWHM_{\rm red} = \gamma \,FWHM_{\rm blue}$. Up to four such components per transition were necessary to recover the observed profiles. In the case of the prominent doublets, we modelled the red and blue components separately with the relative centroid distance tied to the laboratory value and assume equal flux contribution (optically thick case) and broadening. The only exception is \ion{N}{v}, where, due to strong blending with the nearby \ion{Ly}{$\alpha$}, the profile requires one broad asymmetric Gaussian in addition to a somewhat narrower pair of lines of identical profiles for the red and blue transitions to achieve a good fit. No truly narrow components were identified in either \ion{N}{v} or \ion{C}{iv}.

The modelled emission-line parameters are listed in Table \ref{tab:emission}. We emphasise that the model is not intended as a physical characterisation of the emission spectrum, but rather as an empirical description of the emission lines and the continuum, upon which the absorption features are imprinted.

\zw{} is known for its complex, prominent emission of \ion{Fe}{ii} and \ion{Fe}{iii} lines, most notable above 1700 \AA{} \citep[e.g.][]{Vestergaard2001}. In the energy range of the COS spectrum, these lines are densely spaced and blended with other features. While this emission is in principle separable in otherwise continuum-dominated parts of our spectrum, it is difficult to identify any possible contamination in the prominent (non-iron) emission lines. Unfortunately, comprehensive atomic models capable of fully explaining these complex iron features are currently unavailable. Nevertheless, the profiles of our modelled emission lines exhibit no discernible distortions that would point to localised contamination. Therefore, we assume that our results presented below are not significantly impacted.

\subsection{Absorption systems}\label{ssec:absorption}

Several narrow absorption features are evident in the spectra, typically blue-ward of the systemic redshift. To analyse the absorber signatures in different ions, we transformed portions of the spectrum around the emission lines with the identified absorption features into velocity space, relative to the host galaxy cosmological recession velocity. We identified four blueshifted absorption systems detectable in more than one transition and three more features visible only in \ion{Ly}{$\alpha$}.

\subsubsection{Velocity structure}\label{ssec:velocity}

\begin{figure}
    \centering
    \includegraphics[width=1\linewidth]{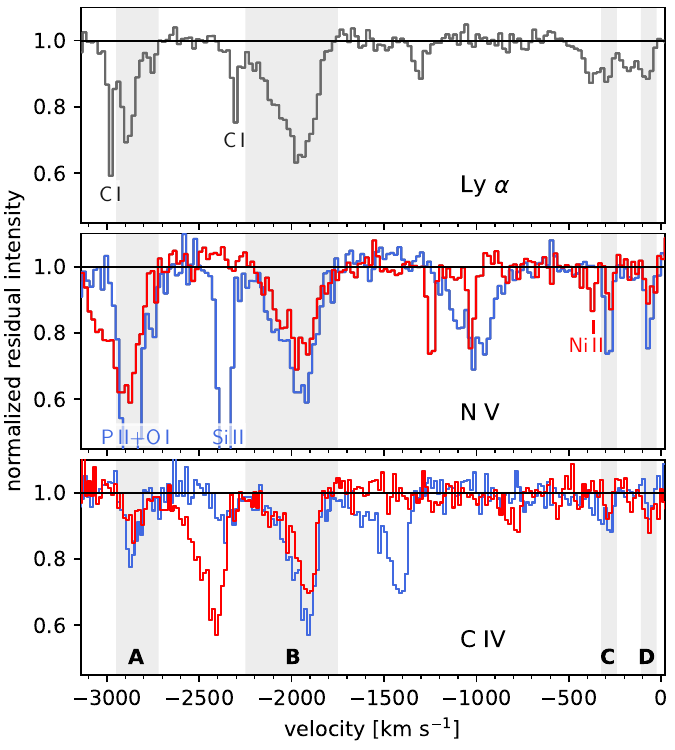}
    \caption{Absorption features in \zw{}. Normalised residual intensities are plotted as a function of velocity relative to the systemic redshift of $z = 0.061169$. Unless identified as ISM lines (labelled), the absorption features are assumed to belong to the displayed line: \ion{Ly}{$\alpha$} $\lambda 1215$ in the top panel, \ion{N}{v} $\lambda 1238$ (blue) and $\lambda 1242$ (red) in the middle panel, and \ion{C}{iv} $\lambda 1548$ (blue) and $\lambda 1550$ (red) in the bottom panel. The four absorption systems that appear in at least two transitions are marked with the shaded areas for clarity and labelled in the bottom panel as A, B, C, and D and discussed in the text.}
    \label{fig:absorption-v}
\end{figure}

The absorption is most prominent in \ion{Ly}{$\alpha$}, \ion{N}{v}, and \ion{C}{iv}. We show the observed spectrum around these ions normalised by the modelled emission in Fig. \ref{fig:absorption-v}. The four absorption systems, indicated with shaded regions, correspond to a mean velocity of $-2900~\rm km\,s^{-1}$, $-1950~\rm km\,s^{-1}$, $-280~\rm km\,s^{-1}$, and $-60~\rm km\,s^{-1}$. In the following, we refer to these systems as A, B, C, and D, respectively.

The highest velocity system, A, can be well identified in \ion{Ly}{$\alpha$} and \ion{C}{iv}. In \ion{N}{v}, the position of the blue component of the doublet coincides with a pair of strong interstellar lines, \ion{P}{ii} and \ion{O}{i}. Only a narrow feature red-ward of the blend is visible. The red component is also blended, this time with the blue component of the strongest ionised absorber at $-1950~\rm km\,s^{-1}$ (component B). This alignment of the two ionised absorbers results from their velocity separation being the same as the \ion{N}{v} $\lambda\lambda1238.83,~1242.80$ doublet component separation ($964~\rm km\,s^{-1}$). While this effect may be, in principle, coincidental \citep[although unlikely, see][]{Ganguly2003}, line-locking is a phenomenon commonly seen in quasar spectra and is interpreted as direct evidence of a line-driven wind. We address the implications of this effect in Sect. \ref{ssec:line-locking}. Weak absorption features at this velocity shift are detected also in \ion{Si}{IV} $\lambda\lambda1393.76,~1402.77$.

Kinematic component B is the deepest one, identified also in the FOS spectrum \citep{Laor1997}. The complex absorption trough covers velocities from $-2230~\rm km\,s^{-1}$ to $-1750~\rm km\,s^{-1}$ relative to the systemic redshift and shows an asymmetric profile, with the deepest absorption around $-1950~\rm km\,s^{-1}$. Besides the features shown in Fig. \ref{fig:absorption-v}, absorption at the corresponding velocity is found also in the \ion{P}{V} $\lambda\lambda1117.98,~1128.01$ doublet and in \ion{Si}{IV}, although significantly weaker.

The last two systems, C and D, are detected close to the \zw{} systemic redshift ($ z = 0.061169 $), at $-280~\rm km\,s^{-1}$ and $-60~\rm km\,s^{-1}$. The absorption is apparent in \ion{N}{v}, red-ward of the emission peak, and is visible also in \ion{Ly}{$\alpha$}, \ion{C}{IV} and in the case of system C also in \ion{Si}{IV}.

In \ion{Ly}{$\alpha$}, we detect three more instances of narrow absorption, at $-1300~\rm km\,s^{-1}$, $-370~\rm km\,s^{-1}$, and $-170~\rm km\,s^{-1}$. However, these do not have any detectable counterparts in other transitions, and thus we cannot conclusively connect them to the AGN environment, in contrast with the absorption systems described above.  

\subsubsection{Column densities and ionisation}\label{ssec:coldens}

\begin{figure}
    \centering
    \includegraphics[width=1\linewidth]{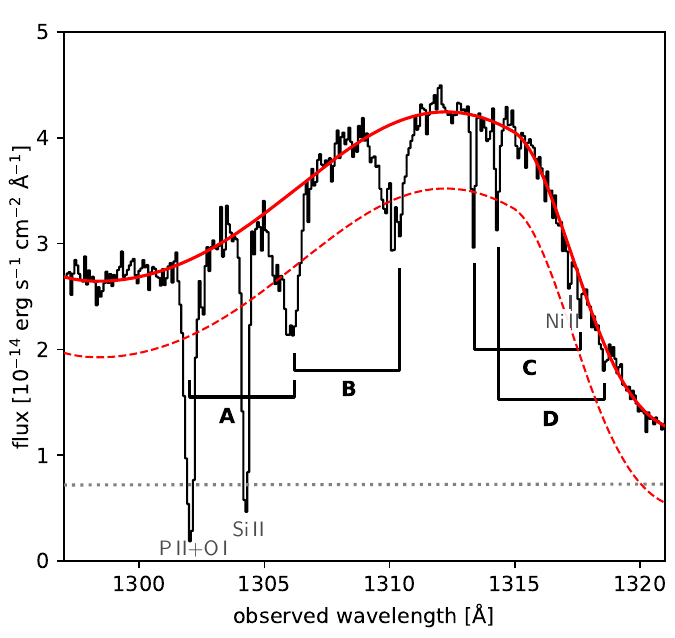}
    \caption{HST-COS view of the \zw{} absorption-affected \ion{N}{v} emission feature. The continuum level is depicted as a grey dotted line, the emission line profile as a red dashed line and the total modelled emission is shown as a solid red line. The positions of \ion{N}{v} absorption features (A, B, C, D) are labelled, as well as the identified ISM lines.
    The strong \ion{P}{ii}+\ion{O}{i} blend overlaps the blue component of the most blue-shifted \ion{N}{v} absorption system.}
    \label{fig:NVline}
\end{figure}

Before proceeding to a detailed investigation of the absorbing gas properties, it is important to assess which components of the emission spectrum are affected by absorption. In Fig. \ref{fig:NVline}, we show a part of the spectrum around the \ion{N}{v} doublet. In absorbers B, C, and D, it is clearly visible that the absorbed flux drops below the continuum-subtracted model. As a consequence, it is reasonable to believe that at least a part of the line-emitting region is absorbed. In the case of absorber A, both doublet components are blended with other features, but the same conclusion can be made from the behaviour in \ion{Ly}{$\alpha$}. As the absorption is rather strong and the continuum source is likely considerably smaller than the line-emitting region, we consider the continuum to be absorbed as well.

\begin{figure*}
    \centering
    \includegraphics[width=0.95\linewidth]{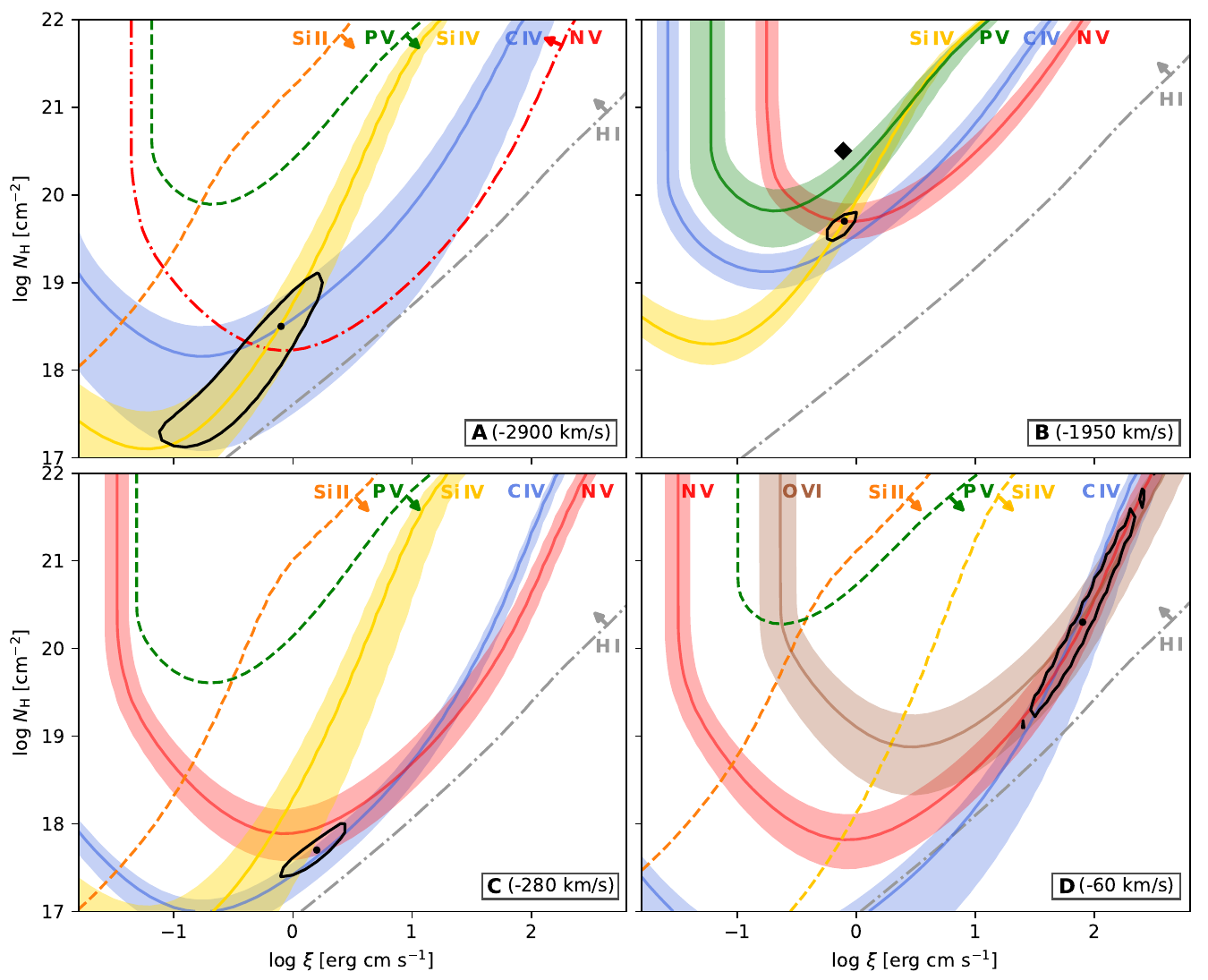}
    \caption{Photoionisation model constraints on the UV absorbers in \zw{}. The measured column densities and the associated uncertainties of each ion are represented by the coloured bands, with the mean value depicted with the solid line. 3$\sigma$ upper limits for undetected lines are shown as dashed lines. The dash-dotted lines, representing the measured column densities of \ion{H}{i} and \ion{N}{v} (in the upper left panel), are regarded as lower limits. For these limits, the permitted portion of the parameter space is indicated with an arrow. In each panel, the black dot depicts the minimal $\chi^2$ value, surrounded by a 1$\sigma$ contour. In the upper right panel (absorber B), the location of the X-ray warm absorber seen at the same radial velocity is represented by a diamond marker, with its size covering the associated 1$\sigma$ uncertainties.}
    \label{fig:Nxi}
\end{figure*}

To characterise the gas properties, we first need to constrain the covering fractions and optical depths of the observed absorption features. To do so, we modelled them in the normalised spectra with negative Gaussian line profiles. The narrow and symmetric absorption lines associated with systems C and D can be described with a single component each. For the wider and deeper A and B, up to four components are necessary to achieve a good description of the data. The best-fitting properties of the modelled components are listed in Table \ref{tab:absorption}.

To mitigate the effects of noise and uncertainties associated with the emission profile modelling, we applied the following simplifying assumptions. We required the doublet components to have the same widths and fix their relative separations to the laboratory values. Furthermore, we required the minimum broadening to be $30~\rm km\,s^{-1}$. For each of the doublets in our data, the oscillator strength of the blue transition is exactly twice that of the red one. Implicitly, the true optical depths of the corresponding lines are also tied by this ratio -- if the small difference in wavelengths is neglected -- given that the doublet components trace the same physical conditions, namely the ionic column density \citep[see][ or eq. (3) below]{SavageSembach1991}. Therefore, we restricted the blue:red line ratios to lie between 2:1 (optically thin case) and 1:1 (optically thick). Finally, given a velocity component, we required this ratio to be the same for all components of each absorption trough.

The contamination of absorber A in \ion{N}{v} requires special treatment. Due to the blending of the red component of system A with the blue component of B, the blue-to-red ratio of system B cannot be independently constrained. Therefore, we fixed this ratio to the value yielded from the fit of the \ion{C}{iv} doublet. For the red component of A, we further fixed the positions of the Gaussian lines that form the profile of this trough to the same velocity shifts as determined from the fit in \ion{Ly}{$\alpha$}, in which the absorption profile is better constrained, and fixed the widths to the same velocity dispersion. However, leaving the normalisations of the modelled troughs free, the resulting strength of the A feature is consistent with zero. The contribution of the blue component in the effectively black \ion{P}{ii}+\ion{O}{i} blend cannot be recovered either, and thus we treat the neighbouring narrow feature, which lies at the low-velocity end of the trough, as marked in Fig. \ref{fig:absorption-v}, and is also clearly visible in \ion{Ly}{$\alpha$}, as a lower limit of the true absorbing column.

From the doublet profiles shown in Fig \ref{fig:absorption-v}, it is clear that at least in the case of systems A and B the blue-to-red line ratio is smaller than 2, but the residual intensities are still relatively high. This indicates that the absorber is only partially covering the central source. To derive the covering fraction, we used the modelled normalised, velocity-dependent, residual intensities $I_1$ and $I_2$ of the red and blue doublet components, respectively. Following \citep{Arav2002}, we used 

\begin{equation}\label{eq:Cf}
    C_\mathrm{f}(v) = \frac{I_1(v)^2 - 2I_1(v) + 1}{I_2(v) - 2I_1(v) + 1},
\end{equation}

\noindent with the assumption that the ratio of the real (unobserved) optical depths of the blue and red components is exactly two. Similarly, we recovered the covering-fraction-corrected optical depths as
\begin{equation}\label{eq:tau}
    \tau_1(v) = -\ln \left( \frac{I_1(v) - [1- C_\mathrm{f}(v)]}{C_\mathrm{f}(v)} \right) = -\ln \left( \frac{I_1(v) - I_2(v)}{1 - I_1(v)} \right),
\end{equation}

\noindent where $\tau_1$ is the optical depth of the red component. We obtained the blue one as $\tau_2 = 2\tau_1$. We note that this velocity-resolved approach has limited effect due to assumptions made during the absorption profile modelling, required by the data quality.

By integrating over the absorption velocity profile, we obtained the ionic column density as

\begin{equation}\label{eq:Nion}
    N_\mathrm{ion} = \frac{m_\mathrm{e}c}{\uppi e^2 f \lambda_0} \int \tau (v) \dif v,
\end{equation}

\noindent where $m_\mathrm{e}$ is the electron mass, $c$ is the velocity of light, $e$ is the elementary charge, $f$ the oscillator strength and $\lambda_0$ represents the laboratory wavelength of the transition \citep{SavageSembach1991}. The use of this relation for the resonance transitions in our data requires the ions to be in their ground states, which we confirm from level population calculations with \textsc{pion} in SPEX \citep[see][]{Mao2017}.

We applied this approach to all identified doublets. The covering fraction at the deepest point of the absorption trough of systems A and B derived from \ion{C}{iv} is $0.45\pm 0.04$ and $0.44\pm 0.04$, respectively. For C and D, the line ratios are consistent with unity covering fractions, but the uncertainties are large and permit also significantly smaller covering of the source. As it is not possible to independently constrain the covering fraction associated with \ion{Ly}{$\alpha$}, we treat the apparent column densities derived for full source coverage as lower limits \citep{Arav2005}. The resulting constraints are listed in Table \ref{tab:ioncoldens}.

Additionally, we place upper limits on column densities of \ion{Si}{ii}~$\lambda 1260.42$, undetected at any of the expected velocity shifts. For each of the four absorption systems, we took the best-fitting profiles from \ion{C}{iv} and fit the normalised spectra at the corresponding wavelengths with only the overall normalisation left free, unless contamination by interstellar lines was apparent. The resulting predicted 3$\sigma$ limits were used to place constraints on the overall properties of the absorbing gas. We assumed the same covering fractions as those derived from \ion{C}{iv}.

Finally, to take advantage of the wavelength coverage down to 910~\AA{} of the archival FUSE data, we examined the longest and most recent exposure, P1110102, taken in November 2000. While molecular hydrogen absorption complicates the identification of most of the ionised lines at the expected blueshifts in this source, the \ion{O}{vi} doublet of component D lies in an apparently absorption-free portion of the spectrum. We analysed this doublet analogously to the COS spectrum features, using one Gaussian profile for each of the lines. The underlying emission was modelled locally by interpolating over the neighbouring absorption-free parts of the spectrum with a polynomial. Assuming that also this absorber is long-lived and variable only on timescales exceeding the separation of the COS and FUSE observations, we included the constraint on the ionic column density in our dataset. The doublet ratio gives $C_\mathrm{f} \sim 0.7$, but this value, as well as the overall strength of the doublet in the spectrum, is dependent on the emission profile representation, constrained with a large uncertainty. That is caused by a large number of neighbouring absorption features and the blended, likely asymmetric, emission of the \ion{O}{vi} doublet. We estimate this additional modelling uncertainty to 0.2~dex of the derived ionic column density.

While a specific column density of an ion can originate in vastly different physical conditions, constraints from several lines can narrow down the range of possible properties of the absorbing gas. To link the measured ionic column densities above to these global characteristics of the gas, namely the hydrogen column density and the ionisation parameter, we used photoionisation models obtained with Cloudy v22.00 \citep{Ferland2017} with the following conditions. For the gas, we assume the turbulent velocity is $30~\rm km~s^{-1}$, consistent with the narrowest features in the data, and the hydrogen particle density is $10^6~\rm cm^{-3}$. We note that the simulation results relevant to this work are not density-dependent for values below $\sim$$10^{11}~\rm cm^{-3}$. Finally, the gas is assumed to be in photoionisation equilibrium with the incident ionising SED. In principle, however, this may not be the case, depending on the recombination timescale of the gas \citep{KrolikKriss1995, Rogantini2022, Gu2023}. However, given the low X-ray continuum variability preceding the HST observations \citep{Wilkins2017}, we expect that any time-dependent effects resulting from a delayed response to the changing incident flux can be neglected relative to measurement errors. 

The ionic column densities are derived for the \zw{} SED detailed in Sect. \ref{ssec:SED} and a grid of ionisation parameter $\xi$ and total column density $N_\mathrm{H}$ with a step of 0.1 dex in each dimension. The ionisation parameter $\xi$ is defined as

\begin{equation}\label{eq:xi}
    \displaystyle \xi = \frac{L_{\mathrm{ion}}}{nr^2}, 
\end{equation}

\noindent where $ L_{\mathrm{ion}}$ is the ionising luminosity between 1--1000\,Ryd \citep{Kallman2001}, $n$ is the hydrogen number density and $r$ is the gas distance from the source of the ionising radiation. We note that for our SED of \zw{}, $\xi$ (in $\rm erg~s^{-1}~cm $) relates to the dimensionless ionisation parameter $U$ as $\log U = \log \xi -1.26 $. This parameter based on \citet{Davidson1977} is defined as the ratio of the incident flux of hydrogen-ionising photons $\phi_\mathrm{H}$ (with energy $>$1~Ryd) to the hydrogen density of the gas, further divided by the speed of light $c$; $U = \phi_{\rm H}/(n c)$.

Fig. \ref{fig:Nxi} shows the photoionisation model predictions for all four absorption systems. The shaded areas specify the 1$\sigma$ regions for conditions in which the measured column densities of each ion are expected. Upper limits provided by non-detections and lower limits derived for \ion{H}{i} and \ion{N}{v} in the case of the highest velocity systems are given as well, to provide more insight into the physical properties of the absorbing gas.

To identify the most likely photoionisation model, we used the \citet{Borguet2012b} definition of the merit function, suitable for the logarithmic values of $N_i$,

\begin{equation}\label{eq:uncertainty}
    \chi^2 = \displaystyle\sum_{i} \left( \frac{\log(N_{i,\mathrm{mod}}) - \log(N_{i,\mathrm{obs}})}{\log(N_{i,\mathrm{obs}}) - \log(N_{i,\mathrm{obs}}\pm \sigma_i)} \right)^2,
\end{equation}

\noindent summing over well-constrained ions. For each ion $i$, $N_{i,\mathrm{obs}}$ and $N_{i,\mathrm{mod}}$ are the observed and modelled column densities, respectively, and the generally asymmetric error on the measured column density $\sigma$ is either added or subtracted, depending on whether the modelled value is smaller or larger than the observed one, respectively. The mean values and the associated uncertainties for all four absorption systems are given in Table \ref{tab:bestfit}.

\begin{table}[]
    \renewcommand{\arraystretch}{1.2}
    \centering
    \caption{Column densities of ions detected in the \zw{} COS spectrum.}
    \begin{tabular}{lcccc}
        \hline\hline
         ion   & \multicolumn{4}{c}{$\log N_\mathrm{ion}~[\rm cm^{-2}]$} \\
               & A & B & C & D \\\hline
        \ion{H}{i} & 13.5$\pm$0.1 \tablefootmark{(a)} & 13.9$_{-0.2}^{+0.1}$\tablefootmark{(a)} & 12.8$\pm$0.1\tablefootmark{(a)} & 12.8$\pm$0.1\tablefootmark{(a)}\\
        \ion{C}{iv} & $14.2_{-1.1}^{+0.3}$ & 15.2 $\pm$ 0.2 & $13.0_{-0.3}^{+0.1}$ & $12.5_{-0.9}^{+0.2}$\\
        \ion{N}{v} &  13.7$_{-0.5}^{+0.3}$\tablefootmark{(b)} & 15.2 $\pm$ 0.2 & 13.4 $\pm$ 0.3 & 13.3 $\pm$ 0.3\\
        \ion{Si}{ii} & <12.8 & $\cdots$ & <11.8 & <12.3\\
        \ion{Si}{iv} & $12.2_{-0.5}^{+0.4}$ & $13.4_{-0.2}^{+0.3}$ & $11.7_{-1.0}^{+0.4}$ & <11.6\\
        \ion{P}{v} & <12.8 & $12.8_{-0.4}^{+0.2}$ & <12.5 & <13.2\\\hline
        \ion{O}{vi} & $\cdots$ & $\cdots$ & $\cdots$ & $15.2_{-0.6}^{+0.4}$\\
    \hline
    \end{tabular}
    \tablefoot{\tablefoottext{a}{Treated as a lower limit due to saturation.}
               \tablefoottext{b}{Treated as a lower limit due to the exclusion of the blended part of the absorption trough.}
                Components A, B, C, and D denote the kinematic absorption systems at $-2900$, $-1950$, $-280$, and $-60~\rm km\,s^{-1}$, respectively. {3$\sigma$ limits are given.} The ionic column densities are given after correcting for partial covering, namely $C_\mathrm{f} = 0.45$ and 0.44 for components A and B, respectively. C and D are given assuming full source coverage. \ion{O}{vi} column density is derived from an archival FUSE spectrum (see the text for more details).}
    \label{tab:ioncoldens}
\end{table}

Given the large number of detected UV features, absorber B has, unsurprisingly, the tightest constraints on its properties: $\log \xi = -0.1\pm0.1$ and $ N_\mathrm{H} = (5\pm2) \times 10^{19}~\mathrm{cm^{-2}}$. For this component, there is a kinematic correspondence with a low-ionisation X-ray warm absorber seen in the RGS spectrum concurrent to the COS spectrum, analysed in detail by \citet{Silva2018}. To compare its properties to the UV absorber, we re-analysed the RGS spectrum in SPEX using the same model with one modification: the SED from Sect. \ref{ssec:SED} was utilised to compute the ionic column densities of the photoionised absorption. The newly obtained ionisation parameter is remarkably similar to the UV constraint, $\log \xi = -0.11^{+0.07}_{-0.06}$, but the column density is higher, $ N_\mathrm{H} = 3.2^{+0.2}_{-0.3}\times 10^{20}~\mathrm{cm^{-2}}$. For completeness, we note the radial velocity of the absorber (modelled with component \texttt{xabs} in SPEX) is constrained from the fit to $ -1940^{+70}_{-130}~\mathrm{km\,s^{-1}}$. The correspondence in ionisation and outflow velocity suggests a direct connection between the X-ray and UV absorption signatures and a common physical origin. If that is the case, the difference in the total column density requires more scrutiny, as it can give more insight into the composition of the absorbing medium. We discuss this further in Sect. \ref{ssec:XUVcolumn}. Finally, from the constraints on the column density and ionisation, we do not expect significant X-ray absorption detectable in the RGS spectrum corresponding to the velocity components A, C, and D.

\begin{table}[]
    \renewcommand{\arraystretch}{1.2}
    \centering
    \caption{Global properties of the UV absorbers.}
    \begin{tabular}{lcccc}
        \hline\hline
         abs.   & $v~\rm [km\,s^{-1}]$ & $C_\mathrm{f}$ & $\log \xi$ & $\log N_\mathrm{H}~[\rm cm^{-2}]$   \\\hline
        A & $-2900$ & $0.45\pm 0.04$ & $-0.1_{-1.0}^{+0.3}$  & $18.5_{-1.4}^{+0.6}$\\
        B & $-1950$ & $0.44\pm 0.04$ & $-0.1 \pm 0.1$        & $19.7_{-0.2}^{+0.1}$\\
        C & $-280$  & 1 & $0.2_{-0.3}^{+0.2}$   & $17.7 \pm 0.3$\\
        D & $-60$   & 1 & $1.9 \pm 0.4$         & $20 \pm 1$\\
    \hline
    \end{tabular}
    \tablefoot{For A and B, the ionisation parameter and hydrogen column density values are derived assuming the covering fraction $C_\mathrm{f}$ derived from doublet component ratios. See the text for more details.}
    \label{tab:bestfit}
\end{table}

%-------------------------------------------------------------------
\section{Discussion}\label{sec:discussion}

\subsection{Dust extinction}\label{ssec:dust}

The UV continuum in \zw{} suggests reddening by a considerable amount of dust in the line of sight. This was pointed out already by \citet{Laor1997}, who noticed the unusually red spectrum in the HST-FOS data. As described in Sect. \ref{ssec:SED}, a custom extinction law was needed to recover a realistic continuum shape. We note that this curve, derived from the continuum shape observed with COS (below 1160~\AA{}), is consistent also with the archival FUSE observation from 2002, extending the band down to 910~\AA{}. Relative to the SMC extinction law, flattenning above 8 eV (1550~\AA{}) is necessary, suggestive of a different dust size distribution, with a relatively smaller fraction of grains with the corresponding size $ \lesssim\!0.25~\mathrm{\mu m} $. This could be due to past brightening of the AGN, or a nuclear starburst which \zw{} is likely undergoing, as suggested by CO emission observations with ALMA, reported in \citet{Fei2023}, and also other works \citep[e.g.][]{Barvainis1989, Eckart1994}, that resulted in the destruction of the smaller grain population \citep{Chang1987, Laor1993}.

In the photoionisation modelling of the UV absorbers (Sect. \ref{ssec:coldens}), we assume the dust is located farther away from the AGN centre. It might be, however, possible that the dust is present more locally and that the absorbing gas is exposed to the extincted SED. In the case of our analysis, we do not get an exact overlap in the column density-ionisation parameter space from the measured ionic column densities (see Fig. \ref{fig:Nxi} and Sect. \ref{ssec:XUVcolumn} below). This could indeed be explained by a difference in the true ionising SED and the one used for the photoionisation modelling. Unfortunately, it is difficult to determine the SED outside of the spectral bands covered with the HST and \textit{XMM-Newton} data. Due to the atypical dust distribution, suggested by the custom extinction curve, it is not possible to place any reliable limits on the amount of the neutral gas from the observed reddening, as opposed to the known relation for the Galactic dust. More importantly, the UV spectrum does not show any significant atomic gas features at the \zw{} redshift, which would point at a considerable column of neutral gas in the line of sight and thus suppress the ionising flux in the EUV band.

The effect the `missing' ionising radiation has on the photoionisation models is examined in \citet{Kara2021}. There, the authors were able to determine the relative position of an ionised absorber and a UV--X-ray obscurer in Mrk~817 by comparing the predictions for ionic concentrations for an obscured and an unobscured SED. In the case of obscuration in the EUV and soft X-ray bands, the effect on the resulting column densities is quite large, as the number of ionising photons changes substantially. The situation in \zw{}, however, differs greatly from that in Mrk~817. The medium responsible for the reddening in our data appears to have a large dust-to-gas ratio. With that, the absorbed SED above the Lyman limit is likely not very different, and thus we do not expect the results to change significantly if the absorbers `see' the reddened continuum.

\subsection{Line-locking}\label{ssec:line-locking}

The kinematic components A and B are detected at the respective outflow velocities of $-2900~\rm km\,s^{-1}$ and $-1950~\rm km\,s^{-1}$, with the difference matching the \ion{N}{v} doublet separation. Unfortunately, detailed analysis of the $-2900~\rm km\,s^{-1}$ \ion{N}{v} absorption is hampered by severe blending of the blue doublet component with interstellar lines. The kinematic correspondence is, however, confirmed by \ion{Ly}{$\alpha$} and \ion{C}{iv} absorption. While it is, in principle, not possible to rule out chance alignment, dynamically important line driving, evidenced by this line-locking effect, would be unsurprising in \zw{}, given its quasar-like properties. For effective line-locking, it is necessary that the shielded absorbing gas has the same or smaller covering fraction than the shielding counterpart. And that is indeed the case of the outflows in \zw{}, where the derived covering fractions are very similar. 

While line-locking in \ion{C}{iv} (with velocity separation $500~\rm km\,s^{-1}$) is more common, the \ion{N}{v} separation of $964~\rm km\,s^{-1}$ has been observed as well, e.g. by \citet{Srianand2002, Ganguly2003, Veilleux2022}. Furthermore, in \zw{}, the acceleration due to absorption by \ion{N}{v} is higher than by \ion{C}{iv}, implying its greater importance in radiative driving. This results from the fact that the radiation pressure force is proportional to the absorbed flux, which is higher in \ion{N}{v} relative to other detected absorption features from different ions, as can be read from Figures \ref{fig:COS} and \ref{fig:absorption-v}. It is important to note that while geometrical alignment of the two outflows along the line of sight is necessary, the physical separation is not constrained, except that the faster absorber A must be farther than absorber B for line-locking to occur.

The shaded regions in Fig. \ref{fig:Nxi} visualise conditions in which column densities of different ions are at the observed values. The tightly constrained position of the shielding absorber B in this parameter space allows us to speculate why line-locking in \ion{N}{v} can be successfully maintained in this outflow. When the ionisation parameter changes, as a result of the ionising source variability, the strength of the \ion{Si}{iv} or \ion{C}{iv} absorption lines will change significantly. This translates in Fig. \ref{fig:Nxi} in that the predicted column density of these ions will move away from the shaded area already with a relatively small deviation of $\log \xi$. In the case of \ion{N}{v}, the sensitivity is much smaller, as for the found total column density the absorption troughs will stay similarly deep if $\xi$ changes within 1~dex from the 2015 value. We caution that while this factor alone plays in favour of \ion{N}{v} as the more likely line-locking ion with respect to similarly strong and more often witnessed \ion{C}{iv}, the actual dynamical conditions depend also on the covering fraction across the velocity profile of the shielding trough and other effects that can contribute significantly to the presently observed line-locking.

We find that the contribution of the red \ion{N}{v} doublet component from absorber A to the trough of blue \ion{N}{v} component of absorber B is consistent with zero (see Sect. \ref{ssec:coldens}), which is also consistent with the line-locking scenario. Because if the lack of radiation at the red \ion{N}{v} transition energy prevents further acceleration, the associated absorption signature should be proportionally weaker or even absent, despite the otherwise favourable ionisation state of the gas. If, on the contrary, the line-locking was caused purely by geometrical alignment along the line of sight, the observed absorption of the overlapping components would appear deeper.

In this line-driving scenario, the weak feature clearly detected in \ion{Ly}{$\alpha$} and \ion{N}{v} at $-2755~\rm km\,s^{-1}$ (the low-velocity end of system A troughs) deserves more attention. The red counterpart of the \ion{N}{v} doublet does not fall in the absorption shadow of the kinematic system B. However, if the acceleration via \ion{N}{v} has similar importance as for the absorption trough A, this absorber could be accelerating and might eventually reach the same velocity, merging with the main trough in the spectrum. Evidence for such acceleration, in a line-driven quasar outflow, has been observed by \citet{Hall2007}, based on a displacement of absorption features between observations separated by only 1.4 years. To test for any line shift in our data, we examined an archival COS spectrum (taken between 1150 and 1470~\AA) from January 2012. Despite a lower signal-to-noise ratio of the spectrum, the data quality allowed us to confirm the presence and similar strength of the same absorption systems that we identified in the 2015 spectrum. The location of the $-2755\,\mathrm{km\,s^{-1}}$ feature in \ion{Ly}{$\alpha$} and the blue component of \ion{N}{V} is consistent with the 2015 position within the calibration uncertainty, although an indication of a small (0.04~\AA{}) blue-ward shift in the 2015 spectrum is visible in both lines. Based on this value, we place an upper limit on the radial acceleration, $a \lesssim -10~\rm km\,s^{-1}\,yr^{-1}$. 

It is important to keep in mind that line-locking identifies radiation-driving as an important outward-acting acceleration mechanism, not as the dominant one. In fact, the extended blue tail of the absorption trough of outflow B (see Fig. \ref{fig:absorption-v}) resembles rather the effects of a magnetically-driven wind, presented by \citet{Fukumura2022}. Ultimately, however, the observed absorption profile is determined also by the velocity-dependent absorption measure distribution and other factors, making it difficult to discern if there are other mechanisms at play. 

\subsection{X-ray--UV correspondence}\label{ssec:XUVcolumn}

As described in Sect. \ref{ssec:coldens}, we obtained tight constraints on the column density and the ionisation parameter that best reproduce the observed ionic column densities of the long-seen outflow at $-1950~\rm km\,s^{-1}$. The ionisation parameter gives a remarkably good match with the X-ray warm absorber at the same outflow velocity, which suggests that the UV and X-ray absorption are produced by the same medium. Such UV--X-ray connection has been found in other AGN outflows \citep[e.g.][]{Mathur1995, Gabel2003, Steenbrugge2003, Crenshaw2012}. However, the UV column density in \zw{} derived from our data is by 0.8~dex lower than that of the X-ray gas. 

For \zw{}, the derived ionic column densities of \ion{C}{iv}, \ion{N}{v}, and \ion{Si}{iv} agree on the total column density of $ N_\mathrm{H} \approx 5\times 10^{19}~\mathrm{cm^{-2}}$. The \ion{P}{v} ionic column density suggests a somewhat higher value for the total column, bridging the gap between the X-ray constraint and other ions in the UV. We note, however, that our assumptions about the chemical composition of the gas can reflect in this prediction more than in those from other examined elements. Phosphorus is a weakly abundant element and a small deviation from the assumed solar abundance can have a large impact on the predicted total column density. Elevated phosphorus content in quasar outflows has been reported by e.g. \citet{Arav2001, Borguet2012a}, with the measured values -- corrected for saturation -- reaching ten and four times solar abundance, respectively. A similar situation could apply to our data. A four times higher phosphorus abundance would not only match the combination of the total column density and the ionisation parameter predicted from other lines but could also help explain a rather strong emission of the phosphorus doublet. We note that the apparently stronger emission from the red doublet component (see Table \ref{tab:emission}) is unphysical and may hint at contamination, possibly by \ion{Fe}{iii$^*$} \citep{Ekberg1993}.

The partial covering method used in our analysis allows for a reasonably accurate measurement of the ionic column densities. Nevertheless, the UV column density might still be somewhat underestimated for the following reasons. We modelled the emission spectrum assuming all absorption features are narrow and well-identified, and that no significant blends by e.g. \ion{Fe}{ii}, \ion{Fe}{iii} influence the emission line profiles (see Sect. \ref{ssec:emission}). Nevertheless, a selective contribution to the flux at the position of the \ion{C}{iv}$\lambda 1548.19$ absorption could lower the apparent level of saturation and consequently lead to an underprediction of the total column density. Other doublets seen in absorption are either contaminated by other absorption features or too weak to independently constrain the line ratio. Also, the absorption troughs consist of several components in velocity space, spanning several km~s$^{-1}$ in total. It is therefore possible that the line ratio also varies as a function of velocity, as shown e.g. in \citet{Arav2002, Arav2003}. This effect, however, cannot be isolated in our data due to the modelling uncertainties. Additionally, we constrained the covering fraction velocity profiles in the analysis (see Sect. \ref{sec:analysis-results}), and assumed the absorber is covering homogeneously and uniformly both the continuum and the emission lines, which might not be an accurate representation of the situation and could contribute to the column density discrepancy \citep{Ganguly1999, Gabel2003, Arav2005, Arav2008}.
Finally, it is important to note that the different X-ray and UV column densities can be explained simply by geometric effects, resulting from different wavelength-dependent apparent sizes of both the emitting and absorbing components of the system.

Another apparent disagreement is in the covering fraction. While the covering of the X-ray warm absorber is consistent with unity, the UV-absorbing doublet ratios require a significantly smaller covering fraction, $C_\mathrm{f}\sim 0.4$. Again, an intuitive explanation for such a discrepancy lies in the relative sizes of the underlying, emission-producing regions. The geometrical extent of the X-ray continuum source is considerably smaller than that of the UV-emitting region, and thus while its complete coverage can be in principle easily achieved if the angular size of the absorber is not sufficiently large or if the absorber is clumpy, it might not cover the entire UV source, as observed.

A clumpy structure of the outflow in this source is consistent with the behaviour observed by \citet{Costantini2007, Silva2018, Rogantini2022}, who compare the X-ray warm absorbers in \zw{} observed with RGS in 2002, 2005, 2015, and 2020. They observe variability of both the ionisation parameter and column density, without any apparent relation to the changes in the ionising continuum. Such behaviour, particularly the variable column density, suggests that different parts of the same outflow (at the same radial velocity) are passing across the line of sight in each observation. Beyond the observed motion along the line of sight, an additional transverse velocity component could produce variations in the X-ray column density if the outflow is clumpy. At the same time, the UV absorber could stay relatively stable in its overall properties. Unfortunately, at the other instances when \zw{} was observed with RGS, high-resolution UV spectra, which could shed more light on any variability of the UV properties at these times, are not available.

\begin{figure}
    \centering
    \resizebox{\hsize}{!}{\includegraphics{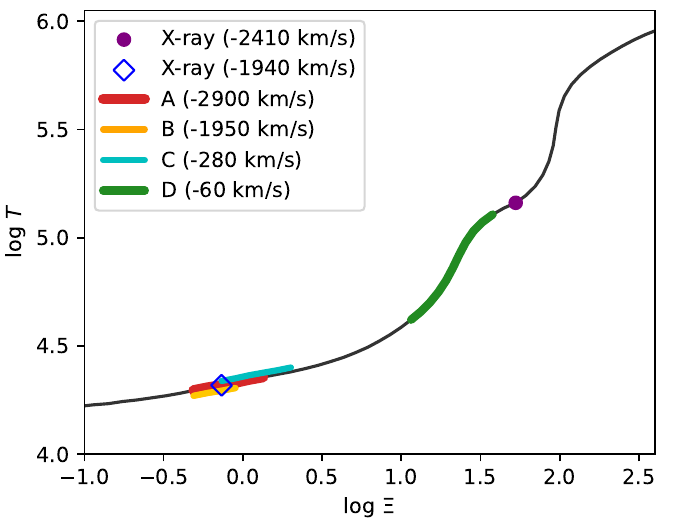}}
    \caption{Thermal stability curve for \zw{}. The ionisation pressure parameter values of the UV and X-ray photoionised absorbers are shown on the curve. For absorber A, the minimal value is obtained as the crossing of the 1$\sigma$ contour in Fig. \ref{fig:Nxi} and the lower limit from \ion{N}{v}.}
    \label{fig:s-curve}
\end{figure}

The velocity and ionisation distribution of the UV and X-ray absorbers provide further evidence for a complex and dynamic environment in the innermost region of the highly accreting object. To provide an overall picture of the ionisation properties of the outflows with respect to the ionising radiation, we visualise the positions of the outflows on the thermal stability curve in Fig. \ref{fig:s-curve}. 
The pressure ionisation parameter $\Xi$ plotted on the $x$-axis is defined as $\Xi = L/(4\uppi r^2 c p) = \xi/(4\uppi c kT)$, where $c$ is the speed of light, $k$ is the Boltzmann constant and $T$ is the electron temperature. The solid black line represents the thermal stability curve, the boundary between the regions with dominant cooling (above the curve) and dominant heating (below) \citep[e.g.][]{Krolik1981}.
It can be seen that the disc-dominated ionising spectrum does not naturally support the development of thermal instabilities in the ionised gas, which is manifested by the absence of the `S' shape of the stability curve, typical for harder SEDs \citep[e.g.][]{Gallo2023}.

\subsection{The emission counterpart of the warm absorber}\label{ssec:abs-em}

\begin{figure}
    \centering
    \includegraphics[width=1\linewidth]{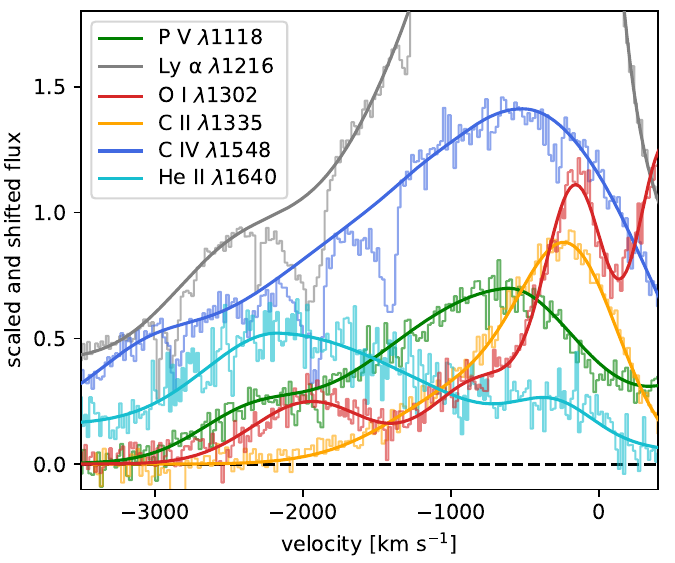}
    \caption{Continuum-subtracted emission lines displayed as a function of velocity relative to the systemic redshift of $z = 0.061169$. The lines are not corrected for blending with nearby features. Besides \ion{Ly}{$\alpha$}, the displayed lines are doublets, with the velocity shift derived for the blue component. Note that the flux increase at approximately $-2300~\rm km\,s^{-1}$ in the \ion{Ly}{$\alpha$} profile coincides with the position of the \ion{Si}{iii}~$\lambda 1206.50$. The flux enhancement at the velocity corresponding to the absorption system B ($-1950~\rm km\,s^{-1}$) in other lines is addressed in the text. \ion{Ly}{$\alpha$} and \ion{C}{iv} are scaled for visualisation purposes.}
    \label{fig:HeII}
\end{figure}

It is possible that the UV spectrum reveals more information about the properties of outflow B. \citet{Laor1997} identified the broad emission feature in the FOS spectrum at $\sim$$1730~$\AA{} in the observer frame as highly blueshifted \ion{He}{ii} emission. Based on the correspondence with the detected ionised absorption velocity shift, the authors argue the emission and absorption could be produced by the same ionised gas, part of which lies in the line of sight. With our higher-resolution HST-COS data, we expand on this idea. In Fig. \ref{fig:HeII}, we show this \ion{He}{ii} emission feature as it appears in our spectrum -- highly asymmetric and double-peaked. The strong, broader component peaks around the velocity shift of $-1950~\rm km\,s^{-1}$, consistent with the strongest absorber (B). The second peak is weaker and narrower and is located at $-300~\rm km\,s^{-1}$, a shift consistent with the location of low-ionisation lines present in the spectrum, such as \ion{C}{ii} or \ion{S}{ii} (Table \ref{tab:emission}). We note that the equivalent width of the stronger \ion{He}{ii} emission line is consistent with quasar scaling relations presented in \citet{Timlin2021}, considering the inherent data scatter. 

We do not conclusively identify any other emission lines shifted to $-1950~\rm km\,s^{-1}$. This is in part due to blending with stronger emission features at the corresponding line shifts and also because of the somewhat uncertain contribution of blended \ion{Fe}{ii} and \ion{Fe}{iii} emission features. Nevertheless, two candidates can be found in the spectrum. An isolated emission line, located at 1373\,\AA, could belong to the \ion{O} {i} $\lambda 1302.17$ transition blueshifted to the correct wavelength. We show also this tentative \ion{O}{i} feature in Fig. \ref{fig:HeII}, along with \ion{Ly}{$\alpha$}, to illustrate the location of the $-1950~\rm km\,s^{-1}$ absorption trough, \ion{C}{iv}, and \ion{P}{v}. The latter also displays enhanced flux around $-2000~\rm km\,s^{-1}$. Interestingly, no increased emission that could be associated with blueshifted \ion{C}{ii} $\lambda \lambda1334.53,~1335.71$ can be seen in the spectrum. 

We stress that the reasoning connecting the B outflow component to a larger, line-emitting, region, is associated with important caveats. Namely, the tentative identification of each of the emission features plotted in Fig. \ref{fig:HeII} is based on their velocity shift with respect to the rest frame of the neighbouring lines. As mentioned in Sect. \ref{ssec:emission}, the line emission from \ion{Fe}{ii} and \ion{Fe}{iii} to the spectrum may provide an alternative explanation for some or all of the observed, apparently strongly blueshifted, features \citep{Wills1985, Laor1997}. 

Nevertheless, with the above-stated caveat, we speculate that if indeed the apparent blueshifted \ion{He}{ii} and \ion{O}{i} emission belongs to the same gas, the absence of other low-ionisation emission lines at this velocity could place limits on the gas properties. Namely, \ion{He}{ii} and \ion{O}{i} are produced primarily in gas of densities $n_\mathrm{e} \sim 10^8\!-\!10^{10}\,\mathrm{cm}^{-3}$. At the same time, the absence of the \ion{C}{ii} doublet is consistent with these constraints, as it would require higher densities (and lower ionisation) to be produced efficiently. Under such conditions, this gas would be more tenuous than the bulk of the broad line region, which is expected to be $n_\mathrm{e} \sim 10^{11}\,\mathrm{cm}^{-3}$ \citep{Laor1997}, and denser than the narrow-line region gas, with $n_\mathrm{e} \sim 10^6\!-\!10^7\,\mathrm{cm}^{-3}$ \citep{Veron-Cetty2004}. Under the assumption that the gas density indeed lies in the range $\sim$$10^8\!-\!10^{10}\,\mathrm{cm}^{-3}$, we can derive a crude estimate of its location directly from the definition of the ionisation parameter $\xi$. Namely, the ionising flux required by the observed $\xi$ places the outflow to $r\sim 10^{17.8}\!-\!10^{18.8}\,\mathrm{cm}$ from the central source. This value range is broadly consistent with the outflow position within the dust sublimation radius, which can be estimated from the bolometric luminosity to $r_{\mathrm{sub}}\sim 10^{18.1}\!-\!10^{18.5}\,\mathrm{cm}$, depending on the dust grain composition \citep{MorNetzer2012}. 

\subsection{Geometrical constraints}\label{ssec:cartoon}

\begin{figure}
    \centering
    \includegraphics[width=0.85\linewidth]{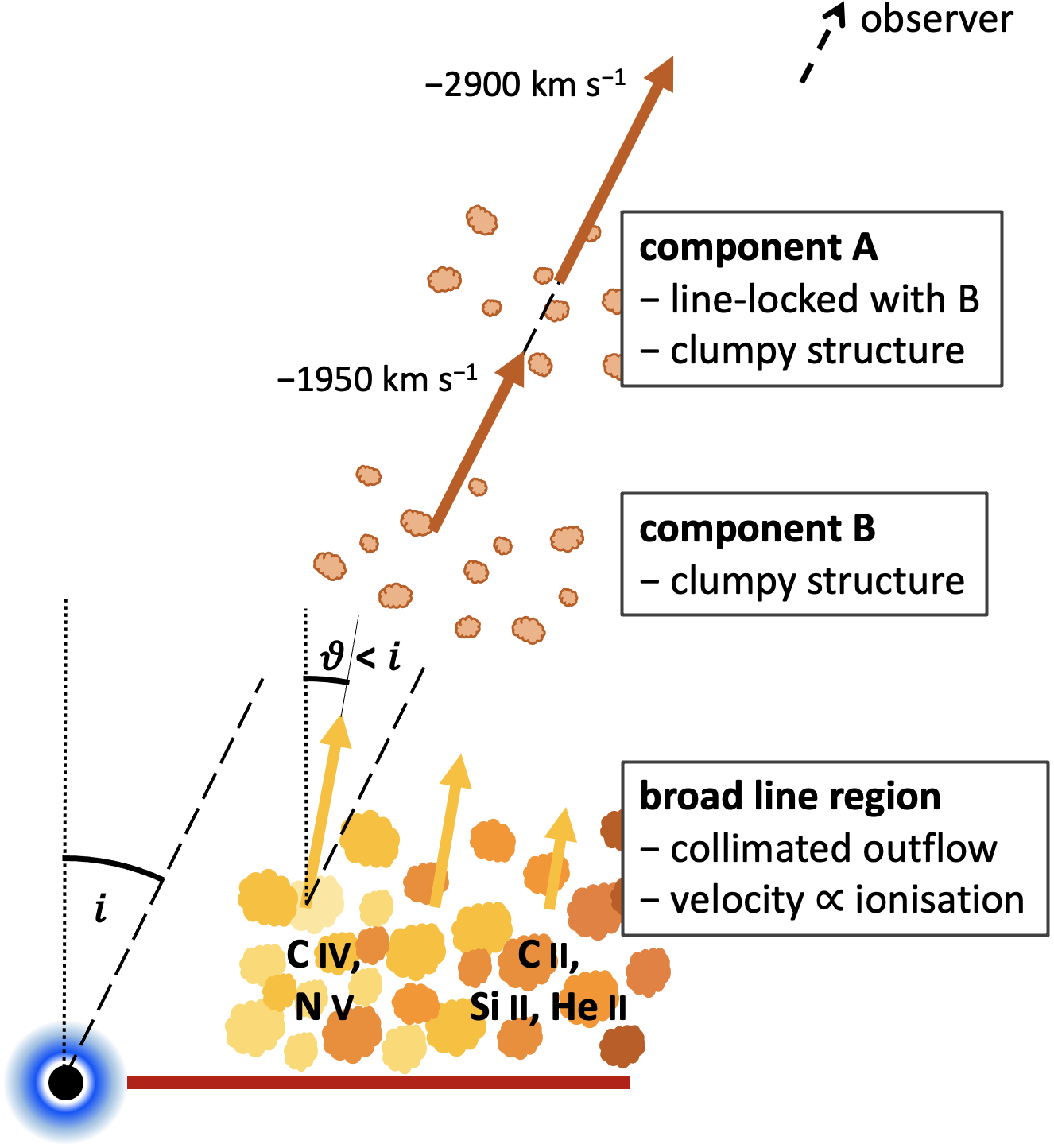}
    \caption{Cartoon of the observed emission and absorption regions in \zw{}. The highly asymmetric profiles with nearly absent red wings can be explained with a rather vertically outflowing BLR. Inwards increasing (radial) velocity of the BLR clouds would result in the detected emission-peak blueshift that grows with ionisation. The outward acceleration of ionised absorption components A and B is directed along the line of sight, which allows \ion{N}{v} line-locking to be seen in the spectrum. Additional transverse motion (e.g. Keplerian, perpendicular to the plane of the image) and a clumpy structure of the outflow B can produce the observed changes in the total column density of the otherwise persistent absorption, detected in the UV and X-ray spectra for over 20 years. If the line emission of unclear origin discussed in Sect. \ref{ssec:abs-em} belongs to blueshifted emission from outflow B, then this gas spans over a larger area and might have originated in the outflowing BLR.}
    \label{fig:cartoon}
\end{figure}

Irrespective of the possible absorption-emission connection in the B component, the rich UV spectrum of \zw{} with its relatively narrow, blueshifted, broad-emission lines as well as absorption features presents valuable insight into the geometry and kinematics of the innermost region of this AGN. We visualise the geometry constraints resulting from our analysis with a cartoon in Fig. \ref{fig:cartoon} and explain our reasoning below.

The highly asymmetric broad emission lines (Fig. \ref{fig:COS} and Sect. \ref{ssec:emission}) with little emission in the red wings suggest that the broad-line producing gas is outflowing at an angle smaller than the AGN inclination \citep[][marked as $\vartheta$ in Fig. \ref{fig:cartoon}]{Murray1997, Chajet2013}, with the receding part obscured by the optically thick accretion disc. Additionally, the increasing blueshift of the emission line peak with increasing ionisation gives evidence for radial stratification in the outflow velocity. 

In Sect. \ref{ssec:coldens}, we show that the four ionised absorption systems detected in the spectrum cover both the emission lines and the continuum, which places the absorbing gas at least as far as the broad line region. The line-locking seen between the kinematic components A and B implies that the outward acceleration of these two absorbers is directed along our line of sight, with A farther away from the ionising source, and that radiative acceleration is strong enough to cause their dynamical coupling. Additionally, the time-dependent behaviour of the X-ray absorption spectrum, discussed in Sect. \ref{ssec:XUVcolumn}, implies a clumpy structure of component B, revealed as a result of sufficient transverse velocity component of the flow. Required by the line-locking and similar UV covering fractions (Sect. \ref{sec:analysis-results}), this property likely applies also to the shielded component A. Finally, due to the weak absorption features associated with components C and D, little can be said about the geometry of these outflows. 

%-----------------------------------------------------------------

\section{Summary}\label{sec:conclusions}
In this paper, we analyse the HST-COS spectrum of \zw{} taken in 2015 as a part of a multi-wavelength campaign. We perform photoionisation modelling of the observed spectral features with the aim of characterising the absorbing medium associated with the AGN. Our major findings from this investigation are the following.
\begin{enumerate}
\item We identify and analyse four kinematic components of ionised absorbers, located at $-60$, $-280$, $-1950$, and $-2900~\rm km\,s^{-1}$ with respect to the cosmological redshift. The corresponding absorption features are detected in \ion{Ly}{$\alpha$}, \ion{C}{iv}, \ion{N}{v}, \ion{Si}{iv}, and in the case of the most prominent absorber also in \ion{P}{v}. See Sect. \ref{ssec:velocity}.
\item For the $-1950$ and $-2900~\rm km\,s^{-1}$ outflows, we constrain the covering fraction from doublet ratios to $0.45\pm 0.04$ and $0.44\pm 0.04$, respectively. The $-60$ and $-280~\rm km\,s^{-1}$ outflows are much weaker, and while their covering fraction cannot be precisely constrained, the \ion{N}{v} line ratios are consistent with optically thin conditions, allowing full covering of the source. See Sect. \ref{ssec:coldens}.
\item The strongest, long-observed outflow, corresponds to an X-ray warm absorber with matching radial velocity of $-1950 \,\mathrm{km\,s}^{-1}$ and ionisation ($\log \xi = -0.1$). The UV column density of this outflow is $ N_\mathrm{H} \approx 5\times 10^{19}~\mathrm{cm^{-2}}$. This value is approximately eight times lower than the one of the X-ray absorber. See Sect. \ref{ssec:coldens} and \ref{ssec:XUVcolumn}.
\item The velocity difference of the $-1950$ and $-2900~\rm km\,s^{-1}$ systems corresponds to the \ion{N}{v} doublet separation, $964~\rm km\,s^{-1}$. The observation of line-locking indicates a coupling between the two kinematic components, suggesting that line absorption plays a significant role in the acceleration of these outflows. Moreover, this effect reveals the direction of outflow motion due to the radiative acceleration, which is aligned with our line of sight. A clumpy structure of the wind combined with sufficient additional transverse motion can explain this geometrical requirement together with long-timescale changes in the X-ray column density of the $-1950~\rm km\,s^{-1}$ absorber reported in the literature. See Sect. \ref{ssec:line-locking} and \ref{ssec:cartoon}. 
\item Based on the observed kinematics, we speculate on a correspondence between the $-1950~\rm km\,s^{-1}$ absorber and an emission component, possibly extended from the broad line region. See Sect. \ref{ssec:abs-em}.
\end{enumerate}

%-------------------------------------------------------------------
\begin{acknowledgements}
    A.J. is grateful to Liyi Gu for useful discussions. W.N.B. acknowledges support from \textit{XMM-Newton} grant 80NSSC22K1806 and the Penn State Eberly Endowment. M.G. is supported by the ``Programa de Atracci\'on de Talento'' of the Comunidad de Madrid, grant number 2022-5A/TIC-24235. We acknowledge financial support from the Space Telescope Science Institute (STScI) Director’s Research Fund during the writing of this paper. Support for Hubble Space Telescope program GO-13811 was provided by National Aeronautics Space Administration (NASA) through a grant from the STScI, which is operated by the Association of Universities for Research in Astronomy, Inc., under NASA contract NAS5-26555. The Space Research Organisation of the Netherlands is financially supported by NWO. This research has made use of the NASA/IPAC Extragalactic Database (NED), which is operated by the Jet Propulsion Laboratory, California Institute of Technology, under contract with NASA. The data analysis and plots presented in this paper can be reproduced using the code and data files available at http://doi.org/10.5281/zenodo.10833894.
\end{acknowledgements}

%-------------------------------------------------------------------
\bibliographystyle{aa}
\bibliography{izw1}

%-------------------------------------------------------------------
\begin{appendix}

\section{Properties of the modelled UV emission and absorption lines}

\begin{table*}[]
    \centering
    \caption{Emission line components properties derived from the 2015 COS spectrum of \zw{}.}
    \begin{tabular}{lccccc}
        \hline\hline
        Feature     & $\lambda_0$ & Flux & $v_\mathrm{cen}$ & FWHM$_\mathrm{b}$ & $\gamma$\tablefootmark{(a)} \\
                    &    \AA    & $10^{-14} \rm erg\,cm^{-2}\,s^{-1}$     & $\mathrm{km\,s^{-1}}$ & $\mathrm{km\,s^{-1}}$ &   \\
        \hline
\ion{P}{V}			 & 	1117.98 & $0.6\pm0.1$ 	 &$-2390\pm40$ 	 &$800\pm100$ 	 &$1.0$ 		\\
\ion{P}{V}			 & 	1117.98 & $6.7\pm0.2$ 	 &$-630\pm20$ 	 &$2100\pm100$ 	 &$0.53\pm0.05$ 	\\
\ion{P}{V}			 & 	1128.01 & $10.3\pm0.1$ 	 &$-630\pm20$ 	 &$2370\pm80$ 	 &$0.50\pm0.03$ 	\\
\ion{C}{III$^*$}		 & 	1175.74 & $3.84\pm0.08$ 	 &$-500\pm20$ 	 &$1430\pm40$ 	 &$0.67\pm0.04$ 	\\
\ion{Si}{II}		 & 	1190.42 & $2.9\pm0.2$ 	 &$-500\pm20$ 	 &$1090\pm50$ 	 &$1.0$ 		\\
\ion{Si}{II}		 & 	1193.29 & $0.9\pm0.1$ 	 &$-250\pm10$ 	 &$460\pm50$ 	 &$1.0$ 		\\
\ion{Si}{II}		 & 	1194.50 & $0.9\pm0.1$ 	 &$-250\pm10$ 	 &$460\pm50$ 	 &$1.0$ 		\\
\ion{Si}{II}		 & 	1197.39 & $0.5\pm0.2$ 	 &$-250\pm10$ 	 &$460\pm50$ 	 &$1.0$ 		\\
unknown    & 	~~~~1273.40\tablefootmark{(b)} & $4.8\pm0.9$ 	 &$\cdots$ 	 &$1300\pm200$ 	 &$1.0$ 		\\
\ion{Si}{III}		 & 	1206.50 & $11\pm2$ 	 &$-200\pm100$  		 &$1050\pm60$ 	 &$1.0$ 		\\
\ion{Ly}{$\alpha$}	 & 	1215.67 & $32\pm3$   &$-390\pm20$ 		 &$680\pm50$ 		 &$1.0$ 		\\
\ion{Ly}{$\alpha$}	 & 	1215.67 & $76\pm6$ 	 &$-600\pm100$ 	 &$2400\pm200$ 	 &$0.7\pm0.1$ 	\\
\ion{Ly}{$\alpha$}	 & 	1215.67 & $49\pm7$ 	 &$-150\pm20$ 	 &$1700\pm100$ 	 &$0.27\pm0.02$ 	\\
\ion{Ly}{$\alpha$}	 & 	1215.67 & $123\pm8$ 	 &$200\pm100$ 		 &$9200\pm400$ 	 &$0.9\pm0.1$ 		\\
\ion{N}{V}			 & 	1238.82 & $9.5\pm0.9$ 	 &$-1100\pm20$ 	 &$2300\pm400$ 	 &$1.0$ 		\\
\ion{N}{V}			 & 	1242.80 & $9.5\pm0.9$ 	 &$-1100\pm20$ 	 &$2300\pm400$ 	 &$1.0$ 		\\
\ion{N}{V}\tablefootmark{(c)}			 & 	1240.80 & $49\pm7$ 	 &$-400\pm40$ 	 &$4200\pm100$ 	 &$0.31\pm0.03$ 	\\
\ion{S}{II}			 & 	1250.58 & $2.0\pm0.2$ 	 &$-270\pm20$ 	 &$1400\pm300$ 	 &$0.4\pm0.1$ 	\\
\ion{S}{II}			 & 	1253.81 & $2.3\pm0.4$ 	 &$-260\pm20$ 	 &$1400\pm300$ 	 &$0.4\pm0.1$ 	\\
\ion{Si}{II}		 & 	1260.42 & $4.3\pm0.3$ 	 &$-330\pm20$ 	 &$940\pm70$ 	 &$0.9\pm0.1$ 	\\
\ion{Si}{II$^*$}		 & 	1264.73 & $4.7\pm0.3$ 	 &$-300\pm20$ 	 &$910\pm60$ 	 &$0.85\pm0.07$ 	\\
unknown   & 	~~~~1372.85\tablefootmark{(b)} & $1.6\pm0.1$ 	 &$\cdots$ 	 &$930\pm70$ 	 &$1.0$ 		\\
unknown   & 	~~~~1378.25\tablefootmark{(b)} & $1.9\pm0.4$ 	 &$\cdots$ 	 &$800\pm100$ 	 &$1.0$ 		\\
\ion{O}{I}			 & 	1302.17 & $3.7\pm0.4$ 	 &$-149\pm8$ 	 &$500\pm30$ 	 &$1.0$ 		\\
\ion{Si}{II} + \ion{O}{I}	 & 	1304.86 & $3.2\pm0.2$ 	 &$-203\pm9$ 	 &$400\pm20$ 	 &$1.0$ 		\\
\ion{O}{I}			 & 	1306.03 & $2.3\pm0.2$ 	 &$-149\pm8$ 	 &$370\pm20$ 	 &$1.0$ 		\\
\ion{O}{I}			 & 	1309.28 & $2.34\pm0.08$ 	 &$-190\pm10$ 	 &$690\pm30$ 	 &$1.0$ 		\\
\ion{C}{II} + \ion{C}{II}	 & 	1335.12 & $2.2\pm0.2$ 	 &$-294\pm8$ 	 &$620\pm30$ 	 &$1.0$ 		\\
\ion{C}{II} + \ion{C}{II}	 & 	1335.12 & $4.6\pm0.2$ 	 &$-670\pm20$ 	 &$1330\pm30$ 	 &$1.0$ 		\\
\ion{P}{III$^*$}		 & 	1344.33 & $1.48\pm0.05$ 	 &$-190\pm30$ 	 &$910\pm60$ 	 &$0.60\pm0.09$ 	\\
\ion{Si}{IV} + \ion{O}{IV]}	 & 	1402.77 & $7\pm1$ 	 &$-590\pm70$ 	 &$2100\pm300$ 	 &$1.0$ 		\\
\ion{Si}{IV} + \ion{O}{IV]}	 & 	1393.76 & $9\pm2$ 	 &$-510\pm40$ 	 &$2630\pm60$ 	 &$0.37\pm0.05$ 	\\
\ion{Si}{IV} + \ion{O}{IV]}	 & 	1402.77 & $13\pm2$ 	 &$-600\pm30$ 	 &$3200\pm700$ 	 &$0.33\pm0.09$ 	\\
\ion{Si}{II}		 & 	1526.71 & $2.5\pm0.2$ 	 &$-230\pm30$ 	 &$1070\pm60$ 	 &$1.0$ 		\\
\ion{Si}{II$^*$}	 & 	1533.45 & $4.5\pm0.9$ 	 &$-230\pm40$ 	 &$1070\pm60$ 	 &$1.0$ 		\\
\ion{C}{IV}			 & 	1548.19 & $9.5\pm0.4$ 	 &$-380\pm20$ 	 &$900\pm30$ 	 &$1.0$ 		\\
\ion{C}{IV}			 & 	1550.77 & $9.5\pm0.4$ 	 &$-380\pm20$	 &$900\pm30$ 	 &$1.0$ 		\\
\ion{C}{IV}			 & 	1548.19 & $15.7\pm0.5$ 	 &$-1180\pm50$ 	 &$2700\pm100$ 	 &$0.31\pm0.06$ 	\\
\ion{C}{IV}			 & 	1550.77 & $15.7\pm0.5$ 	 &$-1180\pm50$ 	 &$2700\pm100$ 	 &$0.31\pm0.06$ 	\\
\ion{He}{II}		 & 	1640.42 & $5.6\pm0.4$ 	 &$-2160\pm70$ 	 &$1200\pm100$ 	 &$1.8\pm0.4$ 	\\
\ion{He}{II}		 & 	1640.42 & $0.7\pm0.2$ 	 &$-300\pm50$ 	 &$600\pm100$ 	 &$1.0$ 		\\
\hline
    \end{tabular}
    \label{tab:emission}
    \tablefoot{
    \tablefoottext{a}{The asymmetry parameter $\gamma$ is defined as $\rm FWHM_r = \gamma \,FWHM_b$, where $\rm FWHM_r$ and $\rm FWHM_b$ are the full-width half maximum values for the red and blue wing of the Gaussian line profile.}
    \tablefoottext{b}{Observed wavelength.}
    \tablefoottext{c}{Blended doublet emission component.}
    }
\end{table*}

\begin{table*}[h]
    \centering
    \caption{Absorption line properties derived from the 2015 COS spectrum of \zw{}.}
    \begin{tabular}{lcccc}
        \hline\hline
        Ion & $\lambda_0$ & EW & $v_\mathrm{cen}$ & FWHM \\
                &    \AA    & \AA & $\mathrm{km\,s^{-1}}$ & $\mathrm{km\,s^{-1}}$ \\
        \hline
\ion{P}{V} & 1117.98 & $0.023\pm0.007$ & $-1890\pm8$ & $60\pm20$ \\
\ion{P}{V} & 1128.01 & $0.012\pm0.004$ & $-1890\pm8$ & $60\pm20$ \\

\ion{H}{I} & 1215.67 & $0.129\pm0.006$ & $-2886\pm2$ & $88\pm5$ \\
\ion{H}{I} & 1215.67 & $0.022\pm0.004$ & $-2755\pm4$ & $50\pm10$ \\
\ion{H}{I} & 1215.67 & $0.04\pm0.01$ & $-2240\pm20$ & $130\pm60$ \\
\ion{H}{I} & 1215.67 & $0.11\pm0.05$ & $-2090\pm20$ & $130\pm20$ \\
\ion{H}{I} & 1215.67 & $0.21\pm0.07$ & $-1954\pm9$ & $130\pm20$ \\
\ion{H}{I} & 1215.67 & $0.033\pm0.004$ & $-1875\pm3$ & $60\pm8$ \\
\ion{H}{I} & 1215.67 & $0.033\pm0.003$ & $-1309\pm3$ & $61\pm7$ \\
\ion{H}{I} & 1215.67 & $0.038\pm0.002$ & $-382\pm2$ & $70\pm2$ \\
\ion{H}{I} & 1215.67 & $0.034\pm0.002$ & $-294\pm2$ & $70\pm2$ \\
\ion{H}{I} & 1215.67 & $0.026\pm0.002$ & $-179\pm3$ & $70\pm2$ \\
\ion{H}{I} & 1215.67 & $0.037\pm0.002$ & $-81\pm2$ & $70\pm2$ \\
\ion{H}{I} & 1215.67 & $0.029\pm0.003$ & $106\pm3$ & $65\pm7$ \\

\ion{N}{V} & 1238.82 & $0.038\pm0.008$ & $-2752\pm4$ & $40\pm10$ \\
\ion{N}{V} & 1238.82 & $0.02\pm0.01$ & $-2160\pm20$ & $70\pm30$ \\
\ion{N}{V} & 1238.82 & $0.13\pm0.04$ & $-2060\pm10$ & $110\pm30$ \\
\ion{N}{V} & 1238.82 & $0.02\pm0.01$ & $-1977\pm3$ & $28\pm9$ \\
\ion{N}{V} & 1238.82 & $0.21\pm0.04$ & $-1922\pm9$ & $120\pm20$ \\
\ion{N}{V} & 1238.82 & $0.02\pm0.01$ & $-1790\pm20$ & $70\pm30$ \\
\ion{N}{V} & 1238.82 & $0.051\pm0.004$ & $-283\pm1$ & $35\pm3$ \\
\ion{N}{V} & 1238.82 & $0.044\pm0.004$ & $-67\pm2$ & $36\pm4$ \\
\ion{N}{V} & 1242.80 & $0.012\pm0.006$ & $-2160\pm20$ & $70\pm30$ \\
\ion{N}{V} & 1242.80 & $0.08\pm0.02$ & $-2060\pm10$ & $110\pm30$ \\
\ion{N}{V} & 1242.80 & $0.018\pm0.009$ & $-1977\pm3$ & $28\pm9$ \\
\ion{N}{V} & 1242.80 & $0.15\pm0.03$ & $-1922\pm9$ & $120\pm20$ \\
\ion{N}{V} & 1242.80 & $0.010\pm0.006$ & $-1790\pm20$ & $70\pm30$ \\
\ion{N}{V} & 1242.80 & $0.025\pm0.002$ & $-283\pm1$ & $35\pm3$ \\
\ion{N}{V} & 1242.80 & $0.022\pm0.002$ & $-67\pm2$ & $36\pm4$ \\

\ion{Si}{IV} & 1393.76 & $0.011\pm0.004$ & $-2887\pm6$ & $30.0$ \\
\ion{Si}{IV} & 1393.76 & $0.045\pm0.005$ & $-1907\pm3$ & $46\pm6$ \\
\ion{Si}{IV} & 1393.76 & $0.008\pm0.003$ & $-1837\pm6$ & $30.0$ \\
\ion{Si}{IV} & 1393.76 & $0.005\pm0.003$ & $-280\pm10$ & $30.0$ \\
\ion{Si}{IV} & 1402.77 & $0.006\pm0.002$ & $-2887\pm6$ & $30.0$ \\
\ion{Si}{IV} & 1402.77 & $0.025\pm0.003$ & $-1907\pm3$ & $46\pm6$ \\
\ion{Si}{IV} & 1402.77 & $0.004\pm0.002$ & $-1837\pm6$ & $30.0$ \\
\ion{Si}{IV} & 1402.77 & $0.003\pm0.001$ & $-280\pm10$ & $30.0$ \\

\ion{C}{IV} & 1548.19 & $0.10\pm0.01$ & $-2861\pm4$ & $90\pm10$ \\
\ion{C}{IV} & 1548.19 & $0.044\pm0.008$ & $-2110\pm10$ & $100.0\pm20$ \\
\ion{C}{IV} & 1548.19 & $0.101\pm0.007$ & $-2010\pm10$ & $100.0\pm8$ \\
\ion{C}{IV} & 1548.19 & $0.20\pm0.03$ & $-1914\pm8$ & $90\pm10$ \\
\ion{C}{IV} & 1548.19 & $0.016\pm0.004$ & $-1880\pm8$ & $40\pm10$ \\
\ion{C}{IV} & 1548.19 & $0.046\pm0.007$ & $-293\pm5$ & $70\pm10$ \\
\ion{C}{IV} & 1548.19 & $0.014\pm0.004$ & $-63\pm5$ & $30.0$ \\
\ion{C}{IV} & 1550.77 & $0.057\pm0.009$ & $-2861\pm4$ & $90\pm10$ \\
\ion{C}{IV} & 1550.77 & $0.033\pm0.006$ & $-2110\pm10$ & $100.0\pm20$ \\
\ion{C}{IV} & 1550.77 & $0.076\pm0.006$ & $-2010\pm10$ & $100.0\pm8$ \\
\ion{C}{IV} & 1550.77 & $0.15\pm0.02$ & $-1914\pm8$ & $90\pm10$ \\
\ion{C}{IV} & 1550.77 & $0.012\pm0.003$ & $-1880\pm8$ & $40\pm10$ \\
\ion{C}{IV} & 1550.77 & $0.023\pm0.003$ & $-293\pm5$ & $70\pm10$ \\
\ion{C}{IV} & 1550.77 & $0.007\pm0.002$ & $-63\pm5$ & $30.0$ \\
                \hline
    \end{tabular}
    \label{tab:absorption}
\end{table*}

\end{appendix}

\end{document}